\documentclass[a4paper,11pt]{article}

\pdfoutput=1 

\usepackage{jheppub} 

\font\sm=cmss10 at 4pt
\usepackage{amsmath}
\usepackage{amsfonts}
\usepackage{amssymb}
\usepackage{graphicx}
\usepackage{simplewick}
\usepackage{youngtab}

\def\q{{q}}
\def\qh{\hat{q}}
\def\qb{\bar{q}}
\def\qbh{\hat{\bar{q}}}

\def\tauh{{\hat{\tau}}}
\def\tbh{{\hat{\bar{\tau}}}}

\def\vt{{\tilde{v}}}
\def\vbart{{\tilde{\bar{v}}}}
\def\wt{{\tilde{w}}}
\def\wbart{{\tilde{\bar{w}}}}

\newcommand{\beq}{\begin{equation}}
\newcommand{\eeq}{\end{equation}}
\newcommand{\beqa}{\begin{eqnarray}}
\newcommand{\eeqa}{\end{eqnarray}}
\def\hw{{\rm hw}}
\def\lab{\label}
\def\Tr{{\rm Tr}}
\def\bul{$\bullet$~}
\def\emr{e^{-\rho}}
\def\pb{\overline{\partial}}
\def\phih{\widehat{\Phi}}
\def\ch{\widehat{C}}
\def\Ah{\widehat{A}}

\def\L{\Lambda}

\def\ra{\rangle}
\def\la{\langle}

\def\cz{{\cal{Z}}}
\def\czb{\overline{{\cal{Z}}}}
\def\g{\gamma}

\def\L{\Lambda}
\def\cH{{\cal H}}

\def\Lb{\overline{\Lambda}}
\def\hsl{hs[$\lambda$]}
\def\Hsl{HS[$\lambda$]}

\def\Lbr{\overline{\Lambda}_{\rho}}

\def\p{\partial}

\def\half{{1\over 2}}
\def\rar{\rightarrow}
\def\mw{{\cal{W}}}
\def\ml{{\cal{L}}}
\def\Lc{{\cal L}}

\def\a{\alpha}
\def\b{\beta}
\def\o{\omega}
\def\l{\lambda}
\def\t{\tau}

\def\ep{e^{\rho}}

\def\zbar{\bar{z}}

\def\Ab{\overline{A}}

\def\ab{\overline{a}}
\def\zb{\overline{z}}
\def\taub{\overline{\tau}}

\def\alphab{\overline{\alpha}}

\title{\boldmath Probing higher spin black holes from CFT}

\author[a]{Matthias R.\ Gaberdiel,}
\author[a]{Kewang Jin}
\author[b]{and Eric Perlmutter}

\affiliation[a]{Institut f\"ur Theoretische Physik, ETH Zurich, CH-8093 Z\"urich, Switzerland}
\affiliation[b]{DAMTP, Centre for Mathematical Sciences, University of Cambridge, CB3 0WA, UK}

\emailAdd{gaberdiel@itp.phys.ethz.ch}
\emailAdd{jinke@itp.phys.ethz.ch}
\emailAdd{E.Perlmutter@damtp.cam.ac.uk}

\abstract{In a class of 2D CFTs with higher spin symmetry, we compute thermal two-point functions of certain scalar primary operators in the presence of nonzero chemical potential for higher spin charge. These are shown to agree with the same quantity calculated holographically using scalar fields propagating in a charged black hole background of 3D higher spin gravity. This match serves as further evidence for the duality between ${\cal W}_N$ minimal models at large central charge and 3D higher spin gravity. It also supports a recent prescription for computing boundary correlators of `multi-trace' scalar primary operators in higher spin theories.}

\begin{document} 

\maketitle

\flushbottom

\section{Introduction}

There is by now plenty of evidence supporting conjectures of holographic duality between 3D theories of higher spin gravity and 2D CFTs with higher spin symmetry (see \cite{Gaberdiel:2012uj} for a review). Such dualities are not strong-weak dualities in the traditional AdS/CFT sense: because the CFTs are exactly solvable in principle, one can often compute a given quantity exactly on both sides, akin to the strong form of the AdS/CFT correspondence. In this sense, a primary motivation for the study of higher spin theories --- aside from possible connections to high energy string theory, among others --- is the possibility that we might extract deeper lessons about holography and the emergence of classical spacetime more generally. 

We focus henceforth on the non-supersymmetric duality proposals of \cite{Gaberdiel:2010pz, 1205.2472}, which state that the ${\cal W}_N$ minimal models in certain large central charge limits, parameterized by a constant $\l$, are dual to the bosonic 3D higher spin theories of \cite{hep-th/9806236} with a single complex scalar field. The CFT has a classical $\mw$-symmetry, denoted $\mw_{\infty}[\l]$ \cite{Henneaux:2010xg,Campoleoni:2010zq,1101.2910, 1205.2472}, which emerges as the asymptotic symmetry algebra of AdS$_3$ higher spin gravity \cite{1101.2910}, where $\l$ parameterizes a line of AdS$_3$ vacua.

The bulk theory also admits other solutions, notably BTZ black holes and their higher spin generalizations (see \cite{1208.5182} for a review). Much about their thermodynamics is understood (though see \cite{Campoleoni:2012hp, Perez:2013xi, 1302.0816, Compere:2013gja} for further scrutiny) and has passed a nontrivial holographic test \cite{1108.2567, 1203.0015}, but the nonlocal nature of higher spin gravity and its large symmetry algebra render geometric interpretations hazardous. In fact, the assignment of the name `black hole' is largely motivated by reference to a Wilson loop involving a higher spin gauge field wrapping a contractible cycle, rather than to, say, the existence of an event horizon \cite{Gutperle:2011kf}. It remains a fascinating problem to find satisfying and gauge invariant generalizations of spacetime geometry to the higher spin world.

One way to improve the physical understanding of a solution of higher spin gravity is to examine scalar fluctuations. The higher spin symmetry fixes the dynamics and the masses of the scalar fields in the theory, which is appealing from a physical perspective but makes computations non-trivial. In \cite{1209.4937}, the authors computed the bulk-boundary propagator of a scalar field propagating in the background of the 3D higher spin black hole of \cite{1108.2567}, to first order in the higher spin chemical potential. The black hole solves the bosonic 3D higher spin theory of \cite{hep-th/9806236} which, at linearized level around a solution in the pure gauge sector, couples free scalar matter to an \hsl$\times$\hsl\ Chern-Simons theory of higher spin fields; for technical reasons, \cite{1209.4937} worked at the specific value $\l=\frac{1}{2}$, where there exists a simpler representation of the hs[$\frac{1}{2}$] algebra in terms of harmonic oscillator variables.

By taking the bulk point to an asymptotic boundary, one can extract two-point functions of dual CFT operators in the presence of a higher spin deformation to the CFT action. If the operators are on opposite asymptotic boundaries of the Lorentzian black hole (in the eternal black hole sense \cite{Maldacena:2001kr}), this represents a `mixed' correlator evaluated in an entangled state of two copies of the boundary CFT. The higher spin calculation of \cite{1209.4937} showed that the first order correction to such a mixed scalar two-point function, analytically continued to Lorentzian signature, is nonsingular. This was argued, by analogy with the same property of the BTZ black hole, to lend support to the claim that the background solution is indeed a black hole with causally disconnected boundary components.

If the operators lie on the same boundary, then the two-point function is simply a thermal correlator for a given scalar operator in a single CFT. In our case, it is calculated in the presence of a nonzero higher spin chemical potential. This one-sided correlator, which was also calculated in \cite{1209.4937} to first order, will be our focus of this paper. 

The goal of this work is to extend the gravity calculations of \cite{1209.4937}, and subsequently match them to a CFT calculation. Indeed, where the gravity and CFT results overlap, they will be seen to match, providing evidence for the proposed holographic dualities at finite temperature that goes beyond the higher spin sector alone. There are actually two different large $c$ limits of the ${\cal W}_N$ minimal models for which dualities have been proposed, known as the 't Hooft \cite{Gaberdiel:2010pz} and semiclassical \cite{1205.2472} limits, and our work applies to both of these. 

More specifically, whereas \cite{1209.4937} worked at the specific value $\l=\frac{1}{2}$ and only to leading nontrivial order in the higher spin chemical potential, we will extend those results to other values of $\l$ and beyond linear order. We will utilize various tools at our disposal for computing the boundary two-point function that were developed in \cite{Vasiliev:1992gr, 1106.2580, 1111.3926, 1209.4937, 1302.6113}. 

Turning to CFT, the dual quantity we must calculate is the torus two-point function of a scalar primary $\phi$ (and its conjugate), in the presence of a deformation of the CFT action by a holomorphic spin-3 operator $W$. The operator $\phi$ has conformal dimension $\Delta=1+\l$ (as well as a fixed set of higher spin zero mode eigenvalues), and indeed such an operator lies in the spectrum of the ${\cal W}_N$ minimal models at large $c$. As we discuss in detail, at $n^{\rm th}$ order in perturbation theory, the problem is reduced to integrating correlation functions of $\bar\phi\phi$ with $n$ spin-3 fields over $n$ copies of the torus. Because we are interested specifically in the asymptotically high temperature regime, we can use a modular transformation to relate them to correlators at very low temperatures, which can be extracted from those on the sphere. With sufficient care, this technique allows us to derive the two-point function to arbitrary order for generic $\l$ using OPEs and methods of contour integration; we content ourselves with a second order calculation. These manipulations are similar in spirit to those of \cite{1203.0015} where the higher spin black hole entropy was reproduced from CFT.\footnote{There are differences however, and the application of the present formalism to the derivation of the black hole entropy is subtle. We will comment on the relation between the two approaches. }

Furthermore, we repeat the above calculations in both bulk and boundary for the case where the dual scalar operator is a `multi-trace' operator, applying a recent prescription for computing correlation functions involving such operators in higher spin gravity \cite{1302.6113}. This involves taking the bulk master fields of the Vasiliev theory to lie in a higher representation of \hsl. We can state this most clearly with reference to the ${\cal W}_N$ minimal models, where representations are labeled by two integrable highest weight representations, $(\L^+;\L^-)$, of the $\mathfrak{su}(N)$ affine algebra at level $k$ and $k+1$, respectively \cite{Goddard:1986ee, Bais:1987zk}. The calculations described above use a bulk scalar field with $m^2=-1+\l^2$, dual to the scalar primary $\phi \equiv ({\rm f};0)$ which is the highest weight state of the minimal representation of $\mw_{\infty}[\l]$. But the CFT has many other representations, and those of the form $(\L^+;0)$ can be made by taking tensor products of the basic $({\rm f};0)$ field, hence the nickname `multi-trace' operators. In our calculations we focus specifically on the $({\sm\yng(1,1)}\ ;0)$ operator, where in the bulk we take the master fields to live in the ${\sm\yng(1,1)}$ representation. Once again, the two sides of the calculation agree. 

The paper is organized as follows. Section \ref{sec:bulk1} carries out the gravity calculations, starting with a few new results regarding the higher spin black hole of \cite{1108.2567}. This is followed by a brief collection of techniques one can use to study propagating matter in higher spin gravity, after which we apply them to the calculation of scalar correlators in the black hole background. In Section \ref{sec:cft} we turn to the CFT side, explaining how one computes the relevant torus amplitude and executing the algorithm through second order in the higher spin deformation. The final result is a formula for generic $\l$, which matches the bulk where applicable. Section \ref{sec:higher} repeats the analysis, now for the $({\sm\yng(1,1)}\ ;0)$ operator in the CFT and the corresponding scalar master field in the bulk. In Section \ref{sec:discussion} we conclude with some discussion. Appendices \ref{hsl} and \ref{app:other} contain some details of the bulk and CFT calculations, respectively, and appendix \ref{app:def} considers the derivation of the higher spin black hole partition function \cite{1203.0015} from our CFT approach.

\section{CFT correlators from scalars in the \hsl\ black hole background}
\label{sec:bulk1}

A central result of \cite{1209.4937} was the computation of the bulk-boundary propagator of a scalar field propagating in the background of the 3D higher spin black hole of \cite{1108.2567}, to first order in the higher spin chemical potential, $\a$. This requires knowledge of only the linearized higher spin theory \cite{hep-th/9806236}, which can be cast as a \hsl$\times$\hsl\ Chern-Simons theory coupled to a scalar of $m^2=-1+\l^2$. The scalar was fixed to obey the alternate quantization at the value $\l=\frac{1}{2}$, and thus it is dual to a scalar operator in a $\mw_{\infty}[\frac{1}{2}]$ CFT with conformal dimension $\Delta=\frac{1}{2}$. Extracting the two-point function where both operators lie on the same boundary at asymptotic infinity --- namely, a torus with modular parameter $\t$ parameterized by Euclidean coordinates $(z,\zb)\sim (z+2\pi,\zb+2\pi) \sim (z+2\pi \t,\zb+2\pi \taub)$  --- the result was found to be\footnote{We always leave implicit the sum over images that enforces the periodicity $(z,\zb) \sim (z+2\pi, \zb+2\pi)$.} 
\beq\label{1.1}
\frac{\la \bar\phi(z,\zb)\phi(0,0) \ra} {\la \bar\phi(z,\zb)\phi(0,0) \ra^{(0)}} = 1+{\a \over 16\t^2}
{3\sin{z\over \t}+(2+\cos{z\over \t})({\zb\over \taub}-{z\over \t})
\over\sin^2{z\over 2\t}}+O(\a^2) \ ,
\eeq
where $\la\bar\phi(z,\zb)\phi(0,0) \ra^{(0)}$ is the thermal two-point function in the absence of a higher spin deformation,
\beq
\la\bar\phi(z,\zb)\phi(0,0) \ra^{(0)}=\sqrt{\frac{1}{4\t\taub\sin {z\over 2\t}\sin{\zb\over 2\taub}}} \ .
\eeq

This is the result we would like to generalize in the bulk and then match to a CFT calculation. We begin with a short treatment of the black hole background itself and free scalar dynamics in 3D higher spin gravity. Subsections \ref{2.1} and \ref{2.2} are mostly review but contain an updated treatment of the black hole including some new results.

\subsection{The \hsl\ black hole}\label{2.1}

Let us start with the Chern-Simons equations of motion
\begin{equation}\label{CSeom}
dA + A \wedge \star A = 0 \ , \qquad d\bar{A} + \bar{A} \wedge \star \bar{A} = 0 \ ,
\end{equation}
where $(A,\bar{A})$ are independent elements of the Lie algebra hs$[\lambda]$; the generators of \hsl\ are denoted as $V^s_m$, see appendix~\ref{hsconvections} for our conventions. It turns out to be convenient to choose a gauge \cite{Campoleoni:2010zq} such that the black hole solution takes the form
\begin{eqnarray}\label{CSgaugefields}
A(\rho,z,\bar{z}) &=& b^{-1} a b + b^{-1} d b \\
\bar{A}(\rho,z,\bar{z}) &=& b \bar{a} b^{-1} + b d b^{-1} \ ,
\end{eqnarray}
where $b = e^{\rho V^2_0}$, and the connections $(a,\bar a)$ are constant with $a_{\rho}=\ab_{\rho}=0$. For the \hsl\ black hole solutions found in \cite{1108.2567}, the unbarred connection has components
\beqa\label{pgf}
a_z &=& V^2_{1} -{2\pi\ml\over k}V^2_{-1} -{\pi\mw\over 2k}V^3_{-2} + {\cal U} V^4_{-3}+\cdots\\
a_{\zb}&=& -{\a\over \taub} \left(a_z\star a_z\right)\Big|_{\rm traceless}  \ ,
\eeqa
where $\star$ denotes the lone star product \cite{Pope:1989sr}, and $(\ml,\mw,{\cal U})$ are the stress tensor, spin-3 and spin-4 charges, respectively. The ellipsis represents an infinite series of higher spin charges, and $k$ is the level of the \hsl\ Chern-Simons action. The solution is accompanied by the analogous barred components $\ab_{\zb}$ and $\ab_z$.
In this gauge it is manifest that each of these constant connections is flat, i.e. the equations of motion (\ref{CSeom}) are simply $[a_z,a_{\zb}] = [\ab_z,\ab_{\zb}]=0$.
%
%

The black hole charges, expanded perturbatively in the spin-3 chemical potential $\a$, are given through $O(\a^2)$ as\footnote{We have followed normalization conventions of \cite{1209.4937}, which differ by a simple rescaling compared to \cite{1108.2567,1208.5182}.}
\beqa\lab{charges}
\ml &=& -{k\over 8\pi \t^2} + {k\over 24\pi \t^6}(\l^2-4)\a^2+O(\a^4) \cr
\mw &=& -{k\over 3\pi \t^5}\a+O(\a^3) \\
{\cal U} &=& {7\over 36 \t^8}\a^2+O(\a^4) \ , \notag
\eeqa
along with barred charges given by $\a\rar \alphab$, $\t \rar \taub$. All other charges are zero at this order. At $\l=0,1$, there are non-perturbative conjectures for the charges $(\ml,\mw)$ obtained from a CFT calculation using free-field realizations of the $\mw_{\infty}[\l]$ symmetry \cite{1108.2567}. 

These charges were obtained by demanding that the holonomy of this solution around the thermal cycle, denoted as
\beq
\cH = e^{\o} \ , \qquad \o=2\pi(\t a_z+\taub a_{\zb})
\eeq
be equal to that of the BTZ black hole, $\cH =\cH_{\rm BTZ}$. 
It was shown in \cite{1209.4937} that $\cH_{\rm BTZ}$ is a central element of the group we call \Hsl\ --- that is, it commutes with all elements $V^s_m$ of the algebra \hsl\ --- from which the statement $\cH = \cH_{\rm BTZ}$ follows if the two are related by conjugation by some element $e^X$, with $X \in$ \hsl. Perturbative evidence for this was provided in \cite{1209.4937}, and indeed this is equivalent to demanding that $\Tr(\o^n) = \Tr(\o^n_{\rm BTZ})$ for all $n$ which implicitly defines the charges. Knowledge of the charges is enough to determine the thermal partition function and hence the black hole entropy, and the ensemble thus constructed is guaranteed to obey the first law of thermodynamics; this network of ideas was recently affirmed and clarified in \cite{1302.0816,1302.1583}.

A complementary perspective is provided by considering an infinite-dimensional matrix representation of $\mathfrak{sl}(2)$. The nonzero matrix elements are
\beqa\lab{mat}
(V^2_0)_{jj} &&= {-\l+1\over 2}-j \notag \\
(V^2_1)_{j+1,j} &&= -\sqrt{(-\l-j)j} \\
(V^2_{-1})_{j,j+1} &&=\sqrt{(-\l-j)j} \ , \notag
\eeqa
where $j=1,\ldots,\infty$. 
From these we can construct the defining representation of the full \hsl\ algebra using the enveloping algebra construction \cite{Bergshoeff:1989ns,Pope:1989sr} (see appendix \ref{hsconvections}), and the lone star product is isomorphic to infinite-dimensional matrix multiplication. When $\l=-N$, an ideal forms, and the upper left $N \times N$ block survives as the $N \times N$ representation of $\mathfrak{sl}(N)$.

Now we note that any linear combination of $\mathfrak{sl}(2)$ generators can be diagonalized to a multiple of $V^2_0$
\beq\lab{ma}
\b V^2_1 + \g V^2_{-1} +\delta V^2_0  = S\left(-2i \sqrt{|M|}V^2_0\right)S^{-1}~, ~~ {\rm where} ~~ M= \left(\begin{matrix}
\b & \delta/2 \cr \delta/2 &\g
\end{matrix}\right)
\eeq
for some matrix $S$. For the BTZ, and hence also the \hsl\ higher spin black hole, the thermal holonomy is given by the exponential of such an object, 
\beq
\o_{\rm BTZ} = \pi\left(2\t V^2_1 + {1\over 2\t}V^2_{-1}\right) \ ,
\eeq
and so
\beq
\cH = e^{\o_{\rm BTZ}} =  S \,e^{-2\pi i V^2_0} S^{-1} =e^{-2\pi i V^2_0} \ .
\eeq
The last equality follows because $e^{-2\pi i V^2_0}$ is central; indeed, it is proportional to the (infinite-dimensional) identity,
\beq
\cH= e^{i\pi(1+\l)}\, {\bf 1} \ .
\eeq

Thus it is rather obvious that $\cH$ is central from this matrix perspective, as are other features of the black hole thermal holonomy which were {\it not} already known: for example, there is a periodicity $\cH_{\l} = \cH_{\l+2n}$ for integer $n$. Integer powers $\cH^n$ are trivially elements of the center as well, corresponding to thermal holonomies of black holes with thermal periodicity $(z,\zb)\sim (z+2\pi n \tau,\zb+2\pi n \taub)$. This helps us answer the question of what exactly the center of the group \Hsl\ is. It tells us that at rational $\l=p/q$, the thermal holonomy of the BTZ black hole and its multiply-wound counterparts form a discrete abelian subgroup $\mathbb{Z}_q(\mathbb{Z}_{2q})$ of the center for $p+q$ even(odd), extending a result of \cite{1209.4937}; on the other hand, it proves that for irrational $\l$, the center of \Hsl\ is U(1).\footnote{Incidentally, this last fact is also implied by the results of \cite{1303.0880}. There is further evidence suggesting that the center of \Hsl\ for any non-integer $\l$ is U(1).} 
Furthermore, one can define a trace operation for these matrices \cite{hep-th/9405116},
\beq
\Tr \, X = {1\over -\l}\lim_{N\rar-\l}\sum_{j=1}^N X_{jj} \ .
\eeq
Then one can show, using \eqref{mat} and \eqref{ma}, that generally 
\beq
\Tr\,  e^{\b V^2_1 + \g V^2_{-1} +\delta V^2_0 } =  {1\over \l}{\sin(\l\sqrt{|M|})\over \sin(\sqrt{|M|})}
\eeq
with $M$ defined in \eqref{ma}. One can check this explicitly for $\l=-N$ using matrices, and at $\l=\frac{1}{2}$ using a harmonic oscillator representation of \hsl\ generators \cite{Agarwal:1971wc, Mehta:1977tj}. 

This is a useful result in the general understanding of how to deal with exponentials of \hsl\ valued elements, which appear, for instance, in finite gauge transformations of \hsl\ Chern-Simons theory. One application in the present context is in the creation of a compact generating function for the traces $\Tr(\o_{\rm BTZ}^n)$: by forming the object $\cH_t = e^{t\o}$ and taking derivatives, the equations defining the black hole charges are
\beq
\Tr(\o^n) = {1\over \l}\lim_{t\rar 0}\left(\p_t^n{\sin(\pi \l t)\over \sin(\pi t)}\right)~, ~~ n\in \mathbb{N}
\eeq
up to the choice of an overall normalization of the trace. Still, to find the charges one must solve these equations as in \cite{1108.2567}.


\subsubsection{A zero temperature limit}\label{sec:zero}

The \hsl\ black hole has a useful zero temperature limit, in which we take $\tau_2 \rar \infty$, $\a \rar\infty$ for fixed $\mu = \a/\taub$. All charges vanish, and the resulting connection is
\beqa \lab{chidef}
a &&= V^2_{1}dz-\mu V^3_{2}d\zbar\notag\\
\ab &&= V^2_{-1}d\zbar \ .
\eeqa
This has been referred to as the `chiral deformation' background in \cite{1111.3926,1209.4937}. Its simplicity will allow us to check scalar correlators in the black hole background against independent calculations in the above limit.

\subsection{Scalar fields in higher spin gravity}\label{2.2}

Next we introduce the essential aspects of the machinery for computing scalar bulk-boundary propagators and correlators in 3D higher spin gravity, which have previously been used in various contexts, e.g. \cite{Vasiliev:1992gr, 1106.2580, 1111.3926, 1209.4937, 1302.6113}.

The scalar field and its spacetime derivatives are packaged in a master field of the higher spin theory which we denote $C$. This object is a spacetime zero-form transforming in the twisted adjoint representation of \hsl\ that obeys the following simple equation
\beq\lab{sc}
dC + A\star C - C\star \Ab = 0 \ .
\eeq
The physical scalar field $\Phi$ is the identity component of $C$, which we denote $\Phi \equiv \Tr(C)$. 

The equation of motion \eqref{sc} is deceptively simple: in a generic on-shell background, it can be a challenge to decouple the different components of $C$ and extract a scalar wave equation for $\Phi$. But because the gauge fields can always be written as locally pure gauge, the gauge symmetry of \eqref{sc} allows us to write down its solutions directly in terms of such gauge functions \cite{Vasiliev:1992gr, 1209.4937}. In particular, the scalar master field can be obtained by a gauge transformation from the gauge in which $A=0$ and $dc=0$, where $c$ denotes the master field $C$ in the $A=0$ gauge. This powerful method avoids the need to tediously extract the wave equation in a given background, let alone to solve it. 

More explicitly, connections that are independent of boundary coordinates $(z,\zb)$ are obtained via \eqref{CSgaugefields} from constant $(a,\ab)$, as for the black hole \eqref{pgf}. Then upon introducing the definitions
\beqa\label{lam0}
\L_0 &=& a_{\mu}x^{\mu} \ , \qquad  \L_{\rho} = b^{-1} \star \L_0 \star b \ , \notag\\
\Lb_0 &=& \ab_{\mu}x^{\mu} \ , \qquad \Lb_{\rho} = b \star \Lb_0 \star b^{-1} \ ,
\eeqa
the scalar field is given by
\beq\lab{sca}
\Phi(z,\zb,\rho;0) = e^{\Delta\rho}\, \Tr\left[e^{-\L_{\rho}}\star c \star e^{\Lbr}\right] \ .
\eeq
The parameter $\Delta$ is the conformal dimension of the dual scalar operator, related to the bulk scalar mass in the usual way, $m^2 = \Delta(\Delta-2)$. Specifying to the case where $\Phi$ is the bulk-boundary propagator, it was shown in \cite{1209.4937} that $c$ is a highest weight state of \hsl. This prescription is conjectured to describe the correct generalization of delta-function boundary conditions to the case of a scalar field propagating in an arbitrary higher spin background. 

To extract the boundary two-point functions where both operators live at positive infinity, the AdS/CFT dictionary then directs us to take the large $\rho\rar \infty$ limit of \eqref{sca}, whereby 
\beqa\label{onesided}
\Phi(z,\zb,\rho;0) \approx e^{-\Delta\rho}\, \langle \bar\phi(z,\zb)\phi(0,0)\rangle + \cdots\ , \qquad \rho\rar\infty \ .
\eeqa
In the black hole background, say, one can also consider a `mixed' correlator --- with operators on opposite boundaries of the global Lorentzian spacetime ---  which probes its causal structure; this would be given by the $\rho\rar-\infty$ limit of \eqref{sca} instead. 

Henceforth we focus on the one-sided correlators \eqref{onesided}. These should match those of any holographic dual CFT with $\mw_{\infty}[\l]$ symmetry and a scalar primary of conformal dimension $\Delta$. (Of course, we have in mind the ${\cal W}_N$ minimal models at large central charge.) To evaluate \eqref{sc}, one must choose a representation of \hsl\ in which $(A,\Ab,C)$ live. 
In particular, in \cite{1302.6113} it was argued that -- at least on the level of free fields -- solving \eqref{sc} with master fields in a general representation of \hsl\ computes CFT correlation functions of dual scalar operators living in the same representation of \hsl, which we recall is the wedge algebra of $\mw_{\infty}[\l]$ \cite{1101.2910}. The simplest case is when $C$ lives in the defining representation of \hsl, in which case it contains a single scalar field with $m^2=-1+\l^2$, and the $\star$ product becomes the lone star product. In the context of the ${\cal W}_N$ minimal models, this bulk field $C$ is dual to the $({\rm f};0)$ primary and its \hsl\ descendants.

A useful fact is that the highest weight state $c$ is a projector \cite{1207.6786, 1209.4937, 1303.0880}. In the infinite-dimensional matrix representation of the defining representation of \hsl, it can be written as 
\beq
c = {\rm diag}(1,0,0,\ldots) \ .
\eeq
Then the scalar propagator \eqref{sca} boils down to a single matrix element,
\beq\lab{scb}
\Phi(z,\zb,\rho;0) = e^{\Delta\rho}\la 1|\, e^{\Lbr} e^{-\L_{\rho}} \, |1\ra \ .
\eeq
All of these manipulations are especially simple and transparent when $\l=-N$ is an integer: the master fields become finite-dimensional matrices and the (linearized) bulk theory is $\mathfrak{sl}(N)\times\mathfrak{sl}(N)$ Chern-Simons theory consistently coupled to matter. This case was studied in some detail in \cite{1205.2472, 1210.8452, 1302.6113}. One very useful feature of the $\l=-N$ case is that one can calculate boundary correlators without computing the full propagator. Denoting the highest and lowest weight states of some representation using a bra-ket notation as $|{\rm hw}\ra$ and $|{\rm -hw}\ra$, respectively, the boundary two-point function for operators living in the same CFT is simply
\beq\label{hw}
\langle \bar\phi(z,\zb)\phi(0,0)\rangle = \la {\rm -hw}|e^{-\L_0}|{\rm hw}\ra \,  \la {\rm hw}|e^{\Lb_0}|{\rm -hw}\ra  \ .
\eeq
The large $\rho$ limit eliminates the other terms contributing to \eqref{sca}. 
%
%
\smallskip

It is worth emphasizing that at generic $\l$, generating solutions to scalar wave equations by passage from $A=0$ gauge as in \eqref{sca} is not calculationally feasible in an arbitrary higher spin background. One can overcome these hurdles for integer $\l$, where we have finite-dimensional matrices at our disposal \cite{1302.6113}; at $\l=\frac{1}{2}$, using the harmonic oscillator realization of \hsl\ generators \cite{Agarwal:1971wc, 1209.4937}; or for generic $\l$ if the background is simple enough, by using \eqref{scb} and the matrix representation introduced earlier \cite{1302.6113}. Of course, one can always proceed by using brute force to find the wave equation and solve it. In what follows, we will use each of these methods.

\subsection{Scalar correlators in the black hole background}

We now present our results for scalar correlators in the \hsl\ black hole background, extending \eqref{1.1}. We treat the case of the elementary scalar field in the defining representation of \hsl, with $m^2=-1+\l^2$, dual to a scalar primary operator with $\Delta=1+ \l$. To achieve the alternate quantization, simply take $\l \rar -\l$. 

First, some preliminaries. Because the problem factorizes into barred and unbarred sectors, we will only consider a nonzero left-moving potential, $\a$. At $O(\a^0)$ one has the pure BTZ solution. The full bulk-boundary propagator (up to overall normalization) for the scalar in the background of a rotating BTZ black hole is 
\beq\lab{btzprop}
\Phi^{(0)} = \left({e^{\rho}\over \cos{z\over 2\t}\cos{\zb \over 2\taub}+4e^{2\rho}\t\taub \sin{z\over 2\t}\sin{\zb\over 2\taub}}\right)^{\Delta} \ .
\eeq
One deduces the correct thermal two-point function between (large $\rho$) CFT operators as
\beq\lab{btz}
\la \bar\phi(z,\zb)\phi(0,0)\ra^{(0)} = \left(4\t\taub \sin{z\over 2\t}\sin{\zb\over 2\taub}\right)^{-\Delta}  \ .
\eeq
%
As above, $(z,\zb)$ parameterize the boundary torus.
At higher orders, we expand as
%
\beq\lab{exp}
\la \bar\phi(z,\zb)\phi(0,0)\ra = \la \bar\phi(z,\zb)\phi(0,0)\ra^{(0)}+ \sum_{n=1}^{\infty} \la \bar\phi(z,\zb)\phi(0,0)\ra^{(\a^n)} \ ,
\eeq
where $\la \bar\phi(z,\zb)\phi(0,0)\ra^{(\a^n)}$ is of $O(\a^n)$. We will find it convenient to normalize our results by the leading order piece, so the expansion parameter is the dimensionless ratio ${\a/ \t^2}$.

\subsubsection{First order: Universal structure}

Even at leading non-trivial order in $\a$, it is hard to evaluate \eqref{sca} for the black hole connection \eqref{pgf} at generic $\l$. Instead, we will resort to the brute force solution of the master field equation \eqref{sc}.

Fortunately, this is rather straightforward in perturbation theory. Expand the master fields as
\beq
C = C+ \a\ch\ , \qquad A = A + \a \Ah\ , \qquad  \Ab = \Ab  \ .
\eeq
The $(A,\Ab,C)$ are the leading order BTZ master fields, and at first order we have
\beq\lab{pt1}
d\ch + A\star \ch - \ch\star \Ab = -\Ah \star C \ .
\eeq
The effect of the higher spin terms $\Ah$ is to generate a source term $S\equiv-\Ah \star C$, built out of components of the BTZ master field $C$, for the master field perturbation $\ch$. Our goal is to decouple the components of equation \eqref{pt1} to extract the wave equation for $\phih$.

The source terms can be determined by analyzing \eqref{sc} in the pure BTZ background. It is convenient to expand $S$ in spacetime and along the internal directions, denoting $S^s_{m,x^{\mu}}$ as the component along the generator $V^s_m$ and the differential $dx^{\mu}$. 
%
After some algebra detailed in Appendix \ref{bulkpt}, one arrives at a two-derivative equation for the scalar perturbation $\phih$
\beqa\lab{wave}
&&\left[-3\eta_-^2\p_{\rho}^2 - 6\eta_+\eta_-\p_{\rho} - 768e^{4\rho}\t^2\taub^2(\t^2\p^2 + \taub^2\pb ^2)-192e^{2\rho}\t^2\taub^2\eta_+\p\pb + 3(\l^2-1)\eta_-^2\right]\phih \notag\\
&& \quad = 8e^{\rho}(\l^2-1)\taub^2\eta_-S^2_{1,\zb}+32 e^{3\rho}(\l^2-1)\t^2\taub^2\eta_-S^2_{1,z}\notag \\
&&\qquad  - 768e^{4\rho}\t^2\taub^2(\taub^2\pb S^1_{0,\zb}+\t^2\p S^1_{0,z})-192e^{2\rho}\t^2\taub^2\pb S^1_{0,z} - 3072 e^{6\rho}\t^4\taub^4\p S^1_{0,\zb}  \ ,
\eeqa
where we have defined $\eta_{\pm} \equiv \pm 1 + 16e^{4\rho}\t^2\taub^2$ and explicit expressions of the source terms can be found in (\ref{sources}). One readily confirms that \eqref{btzprop} solves this equation at $S=0$. 

Upon plugging in the source terms computed in the appendix, one is faced with a very long and complicated partial differential equation. Fortunately, we were able to guess the answer. The following solution of \eqref{wave} is the bulk-boundary propagator for the scalar field perturbation:
\beqa\lab{prop}
\phih = {w_{\rm f}\over 2\t^2}\Phi^{(0)}&&\Big[\cos^2{\czb\over 2} \left(\sin\cz+(2-\cos\cz)(\czb-\cz)\right)\notag\\
&&\ -4e^{2\rho}\t\taub \sin\czb\left(2(1-\cos \cz)+\sin\cz \, (\czb-\cz)\right)\notag\\
&&\ +(4e^{2\rho}\t\taub)^2\sin^2{\czb\over 2}\left(3\sin\cz+(2+\cos\cz)(\czb-\cz)\right) \Bigr]\notag\\
&&\times \left(\cos{\cz\over 2}\cos{\czb\over 2}+4e^{2\rho}\t\taub\sin{\cz\over 2}\sin{\czb\over 2}\right)^{-2}  \ ,
\eeqa
where we have defined 
\beq\lab{zdef}
\cz = {z\over \t}~, ~~\czb = {\zb\over \taub}
\eeq
and $w_{\rm f} = \frac{1}{6} (1+\l)(2+\l)$ is the spin-3 zero mode eigenvalue of the scalar field.

Taking the large $\rho$ limit, the $O(\a)$ correction to the boundary two-point function as parameterized by \eqref{exp} is
\beq\label{grav1loop}
{\langle \bar\phi(z,\zb)\phi(0,0)\rangle^{(\a)}\over \langle \bar\phi(z,\zb)\phi(0,0)\rangle^{(0)}}  = {\a \, w_{\rm f} \over \t^2}
{3\sin\cz+(2+\cos\cz)(\czb-\cz)
\over 2\sin^2{\cz\over 2}} \ .
\eeq
This is a nice result. The functional form of both the propagator and the correlator is universal: the only $\l$-dependence enters via the overall constant $w_{\rm f}$ and the mass of the scalar field. We were able to guess this result based on the CFT considerations: as we will justify soon, the universality of the correlator follows from the fact that CFT three-point functions $\la W(x)\bar\phi(z,\zb)\phi(0,0)\ra$ on the plane are fixed by conformal invariance, with an overall coefficient given by the spin-3 eigenvalue of the scalar operator $\bar\phi$, see Section~\ref{sec:first}. For the minimal representation considered here, this eigenvalue is precisely $w_{\rm f}$. The fact that the entire {\it propagator} is universal is less obvious, but true. This is another of the strong restrictions imposed by higher spin symmetry.

\subsubsection{Second order}\lab{a2}

Instead of following the method of the previous subsection to the next order --- where we will almost certainly encounter a partial differential equation whose solution we cannot guess --- we content ourselves with results at discrete, integer values of $\l=-N$. We can jump right to the correlator using \eqref{hw} without solving for the propagator. 

Using an $N\times N$ matrix representation we calculate the following corrections at $O(\a^2)$ and $N=3,4,5,6$, for a scalar in the standard quantization\footnote{Note that $\Delta<0$ in this regime, an aspect of this limit which has been discussed recently in \cite{1205.2472, 1210.8452, 1302.6113, 1303.0880}. The non-unitarity does not come to bear on the calculation of correlation functions.} with $\Delta=1-N$: 
\vskip .1 in

\bul $N=3:$
%
\begin{flalign}\label{n3}
\frac{\langle \bar\phi(z,\zb)\phi(0)\rangle^{(\a^2)}}{\langle \bar\phi(z,\zb)\phi(0)\rangle^{(0)}} = {\a^2\over 36 \t^4 \sin^2 \frac{\cz}{2} }(6+4(\cz-\czb)^2-(6+(\cz-\czb)^2)\cos\cz-6(\cz-\czb)\sin\cz) \ . &&
\end{flalign}
%

\bul $N=4:$
%
\begin{flalign}\label{n4}
\frac{\langle \bar\phi(z,\zb)\phi(0)\rangle^{(\a^2)}}{\langle \bar\phi(z,\zb)\phi(0)\rangle^{(0)}} = {\a^2\over 2\t^4}{(\cz-\czb)^2}  \ . &&
\end{flalign}
%
\vskip .1 in

\bul $N=5:$
%
\begin{flalign}\label{n5}
\frac{\langle \bar\phi(z,\zb)\phi(0)\rangle^{(\a^2)}}{\langle \bar\phi(z,\zb)\phi(0)\rangle^{(0)}} &= \frac{\a^2}{8\t^4\sin^4 \frac{\cz}{2}} \Big[-3+6(\cz-\czb)^2-2(-4+(\cz-\czb)^2)\cos\cz&\notag\\
&\quad +(-5+2(\cz-\czb)^2)\cos2\cz-4(\cz-\czb)(\sin\cz+\sin2\cz)\Big] \ . &
\end{flalign}
%
\vskip .1 in

\bul $N=6:$
%
\begin{flalign}\label{n6}
\frac{\langle \bar\phi(z,\zb)\phi(0)\rangle^{(\a^2)}}{\langle \bar\phi(z,\zb)\phi(0)\rangle^{(0)}} &=\frac{5\a^2}{72 \t^4 \sin^4 {\cz\over 2}} \Big[9(-1+4(\cz-\czb)^2)+8(6+(\cz-\czb)^2)\cos\cz&\notag\\
&\quad +(-39+10(\cz-\czb)^2)\cos2\cz-6(\cz-\czb)(8\sin\cz+5\sin2\cz)\Big]\ . & 
\end{flalign}
%

The functional form of these results is not universal, in parallel with the same property of CFT four-point functions.

\subsubsection{All orders in zero temperature, fixed $\mu$ limit}\label{sec:chiral}

In the limit in which we take the temperature to zero holding $\mu=\a/\taub$ fixed --- introduced in section \ref{sec:zero} --- we can compute the scalar propagator to all orders in $\mu$, for {\it generic} $\l$, by passing from the $A=0$ gauge using \eqref{scb} and the matrix representation \eqref{mat}. For our scalar field $\Phi$ with $\Delta=1+\l$, we thus have
\beq
\lim_{\substack{
  (\t,\taub)\rar\infty, \\
   \mu~ {\rm fixed}
  }}\Phi = e^{(1+\l)\rho}\la 1|e^{e^{\rho}\, \zb V^2_{-1}} e^{-e^{\rho}zV^2_1} e^{\mu e^{2\rho}\, \zb V^3_2}|1\ra \ .
\eeq
The key aspect of the simplicity of \eqref{chidef} is that $V^3_2 = V^2_1 \star V^2_1$. Then using
\beq
\la 1|(V^2_{-1})^p(V^2_1)^q|1\ra   = \delta_{p,q}{q!\Gamma(q+\l+1)\over \Gamma(\l+1)}
\eeq
and expanding in a series, the propagator is
\beqa
\lim_{\substack{
  (\t,\taub)\rar\infty, \\
   \mu~ {\rm fixed}
  }}\Phi &&= e^{(1+\l)\rho}\sum_{m,n,p=0}^{\infty}{(-e^{\rho}z)^m\over m!}{(\mu e^{2\rho}\zb)^n\over n!}{(e^{\rho}\zb)^p\over p!}\la 1|(V^2_{-1})^p(V^2_1)^{m+2n}|1 \ra \\
&&= \left({e^{\rho}\over 1+e^{2\rho}z\zb}\right)^{1+\l}\sum_{n=0}^{\infty}\left[{\mu e^{4\rho}\zb^3\over (1+e^{2\rho}z\zb)^2}\right]^n{\Gamma(2n+1+\l)\over n!\Gamma(1+\l)} \ .
\eeqa
This result matches \cite{1209.4937} at $\l=\frac{1}{2}$ and gives the correct propagator in the $\mu=0$, Poincar\'{e} AdS limit. Taking the large $\rho$ limit, the scalar two-point function is then
\beq\lab{chl}
\lim_{\substack{
  (\t,\taub)\rar\infty, \\
   \mu~ {\rm fixed}
  }}{\la \bar\phi(z,\zb)\phi(0,0)\ra \over \la \bar\phi(z,\zb)\phi(0,0)\ra^{(0)}}= \sum_{n=0}^{\infty}\left({\mu \zb\over z^2}\right)^n{\Gamma(2n+1+\l)\over n!\Gamma(1+\l)} \ .
\eeq
One can confirm agreement with our results through $O(\a^2)$ in this limit, and we will compare \eqref{chl} with a CFT calculation. 

\medskip

Let us now turn to the CFT, where we compute $\langle \bar\phi(z,\zb)\phi(0,0)\rangle$ through $O(\a^2)$ for generic $\l$. Happily, all of the above results will be seen to agree.

\section{The dual CFT point of view}\label{sec:cft}

In \cite{Gutperle:2011kf} the following entry of the higher spin AdS/CFT dictionary was established. 
(We work in Euclidean signature for convenience.) Consider the flat connection, valued in $\mathfrak{sl}(3)$ for simplicity,
\beqa\label{cona}
a &=& \Big(L_1 -{2\pi \over k}\Lc(z,\overline{z}) L_{-1}- {\pi \over 2k} \mw(z,\overline{z}) W_{-2}  \Big) dz
 -\Big( \mu(z,\overline{z}) W_2 + \cdots\Big) d\overline{z} \ ,
\eeqa
where the ellipses represent terms needed to satisfy the equation of motion. With $\mu=0$, this is asymptotic to Poincar\'e AdS in the sense of \cite{Campoleoni:2010zq}, and the dual CFT lives on $\mathbb{R}^2$ parameterized by $(z,\overline{z})$; the equations of motion fix $\ml$ and $\mw$ to be holomorphic currents. For nonzero $\mu$, the CFT action is deformed by a term
\beq\label{actionpert}
\delta S_{\rm CFT} = - \frac{1}{2\pi i} \int dz \, d\bar{z} ~ \mu(z,\overline{z}) \, W(z)
= \frac{1}{\pi} \int d^2 z~ \mu(z,\overline{z})\, W(z) \ , 
\eeq
and one can show that the resulting $\mw_3$ Ward identities are equivalent to the bulk equations of motion. This is the justification for the identification of the $\mu$ in the bulk with the $\mu$ on the boundary, and similarly for the charges $(\ml,\mw)$ with the CFT currents $(L,W)$; indeed, a similar analysis can be done in pure gravity \cite{Banados:2004nr}. This logic extends to the barred sector, and to the replacement of $\mathfrak{sl}(3)$ by any higher spin algebra.\footnote{The authors of \cite{Compere:2013gja, gc} propose a set of boundary conditions which they argue to imply that the $W$-symmetry is preserved even in the black hole background with finite constant $\mu$. They suggest that, correspondingly, the CFT deformation might be viewed as due to an exactly marginal operator. This idea is in apparent tension with the body of results --- including those of the current paper --- that are consistent with the CFT deformation being by an irrelevant operator, though further consideration of their idea is warranted. We thank the authors of \cite{Compere:2013gja, gc} for discussions.}

In this paper, we wish to perturb the BTZ black hole by a chemical potential for the spin-3 charge, cf.~\eqref{pgf}. The CFT, now living on the torus parameterized by $(z,\zb)\sim (z+2\pi,\zb+2\pi) \sim (z+2\pi \t,\zb+2\pi \taub)$, is again perturbed as in \eqref{actionpert} with $\mu$ constant. Our goal is to compute the two-point function of a scalar primary in the presence of this deformation, perturbatively in $\mu$ and in the high temperature regime. As in the bulk calculation, we have turned on a holomorphic chemical potential only. Comparing to the connection \eqref{pgf}, we follow previous work by taking $\mu=\a/\taub$; this will naturally arise from our calculation below (and was already deduced in a different way in \cite{1302.0816,1302.1583}).

For our computations it is actually more convenient to describe the torus as an annulus rather than a parallelogram parameterized by $(z,\zb)$, so let us define the `annulus coordinate', $v=e^{iz}$.\footnote{In what follows, $\lbrace w,w_i,v\rbrace$ parameterize the annulus, while $\lbrace z, z_i\rbrace$ parameterize the parallelogram.} Then transforming \eqref{actionpert} and inserting $\exp[-\delta S_{CFT}]$ into the scalar two-point function, the deformed two-point function is equal to the torus amplitude
\begin{equation}\label{CFTans}
w_1^h \, \bar{w}_1^{\bar{h}} \, w_2^h \, \bar{w}_2^{\bar{h}} \, 
\Tr \Bigl( \bar{\phi}(w_1,\bar{w}_1)\, {\phi}(w_2,\bar{w}_2) \, 
e^{i \frac{\mu}{\pi} \int d^2 v   \frac{v^2}{\bar{v}} W(v) } 
\q^{L_0-\frac{c}{24}} \, \qb^{\bar{L}_0-\frac{c}{24}} \Bigr) \ ,
\end{equation}
where $\q=\exp(2\pi i \tau)$. Using transformation properties of quasi-primary fields, one notes that (\ref{CFTans}) is at least formally doubly-periodic in each $(w_i,\bar{w}_i)$ separately, where the two periodicities refer to $w_i \mapsto e^{2\pi i} w_i$ and $w_i \mapsto \q \, w_i$, respectively. 

Specifying to a primary with  $\Delta=1+\l$, these results should match the bulk calculations of the previous section. In the large $c$ limit of the ${\cal W}_N$ minimal models, the scalar primary $\phi$ is taken to be that of the minimal representation, $\phi \equiv  ({\rm f};0)$, along with its conjugate $\overline{\phi} \equiv  (\bar{\rm f};0)$.\footnote{Note that relative to the conventions of \cite{1101.2910} we have interchanged the roles of $\phi$ and $\bar\phi$.}

We should stress that the exponential in (\ref{CFTans}) differs from what was considered in \cite{1108.2567,1203.0015}, where instead of the 2d integral above a contour integral, leading to $W_0$, was inserted inside the trace. We have explicitly checked\footnote{We thank Per Kraus for discussions and crucial insights on this calculation and the relation between the two approaches.} that one can also derive the scalar correlators in that approach; one must take care to identify properly the evolution operator for the scalars such that the correlator remains periodic. On the other hand, one can imagine computing the black hole entropy using the 2d integral deformation introduced in this paper. As is explained in appendix~\ref{app:def}, this seems to lead to a slightly different result that may however correspond to a different choice of thermodynamic variables \cite{1302.0816}. This issue is however subtle, and we defer its full exploration to the future.\footnote{We thank Tom Hartman for very useful conversations about this issue.}


\subsection{The unperturbed answer}

Let us first explain how to calculate the two-point function at $\mu=0$, i.e.\ the torus two-point 
function
\begin{equation}\label{zero}
F(\bar \phi(w_1,\bar{w}_1) \phi(w_2,\bar{w}_2);\tau, \bar\tau) \equiv
w_1^{h} \, w_2^{h} \, \bar{w}_1^{\bar{h}} \, \bar{w}_2^{\bar{h}} \,
{\rm Tr}\bigl( \bar\phi (w_1,\bar{w}_1) {\phi} (w_2,\bar{w}_2) 
q^{L_0-\frac{c}{24}} \bar{q}^{\bar{L}_0-\frac{c}{24}} \bigr) \ .
\end{equation}
We shall always be interested in scalar fields in this paper, i.e.\ we shall set $\bar{h}=h$ from now on. Since we are interested in this amplitude in the limit $\tau \to 0$, it is advantageous to do a modular $S$-transformation, leading to a trace with modular parameter $\tauh=-1/\tau$. Under the $S$-transformation, the torus amplitude \eqref{zero} transforms as 
\begin{eqnarray}
F \bigl(\bar\phi(w_1,\bar{w}_1) {\phi}(w_2,\bar{w}_2);\tau,\bar{\tau} \bigr) 
= (\tauh \tbh)^{2 h}
F \bigl( \bar\phi (w_1^\tauh,\bar{w}_1^\tbh) {\phi} (w_2^\tauh,\bar{w}_2^\tbh);\tauh,\tbh\bigr) \ .
\end{eqnarray}
In the limit $\tauh_2 \to \infty$, we have $\qh \equiv e^{2 \pi i \tauh} \to 0$, and the leading contribution comes from the vacuum representation, and in fact just from the vacuum state in the trace. Therefore, the zeroth-order term becomes in this limit 
\begin{equation}\label{zerores}
\langle \bar \phi(w_1,\bar{w}_1) \phi(w_2,\bar{w}_2) \rangle^{(0)} 
\cong (\tauh \tbh)^{2 h}  \, 
\frac{w_1^{\tauh h} w_2^{\tauh h} \bar{w}_1^{\tbh h} \bar{w}_2^{\tbh h}}
{(w_1^\tauh - w_2^\tauh)^{2 h} (\bar{w}_1^\tbh - \bar{w}_2^\tbh)^{2 h}} \ ,
\end{equation}
where we have normalized the two-point function by dividing by the zero-point function (which removes the $(\qh \qbh)^{-\frac{c}{24}}$ factor). To compare directly with the bulk result we pass back to the parallelogram via
\begin{equation}\label{plcyl}
w_1 \equiv e^{i z_1} \ , \quad \bar{w}_1 \equiv e^{-i\bar{z}_1} \ , \quad 
w_2 \equiv e^{i z_2} \ , \quad \bar{w}_2 \equiv e^{-i\bar{z}_2} \ ,
\end{equation}
and the zeroth-order term becomes
\begin{equation}\label{2pt}
\langle \bar\phi(z_1,\bar{z}_1) {\phi}(z_2,\bar{z}_2) \rangle^{(0)} = 
\frac{(\tauh \tbh)^{2 h}}{\left( 4 \sin \frac{\tauh (z_1-z_2)}{2}
\sin \frac{\tbh (\bar{z}_1-\bar{z}_2)}{2} \right)^{2 h} } \ .
\end{equation}
This agrees with the gravity result \eqref{btz} after setting $z_1 = z$, $z_2 = 0$, and taking into account that 
\begin{equation}\label{stransform}
\hat{\tau} = - \frac{1}{\tau} \ , \qquad \hat{\bar{\tau}} = - \frac{1}{\bar{\tau}} \ .
\end{equation}

\subsection{First order correction}\label{sec:forder}

\begin{figure}[ht]
\centering
\includegraphics[width=.75\textwidth]{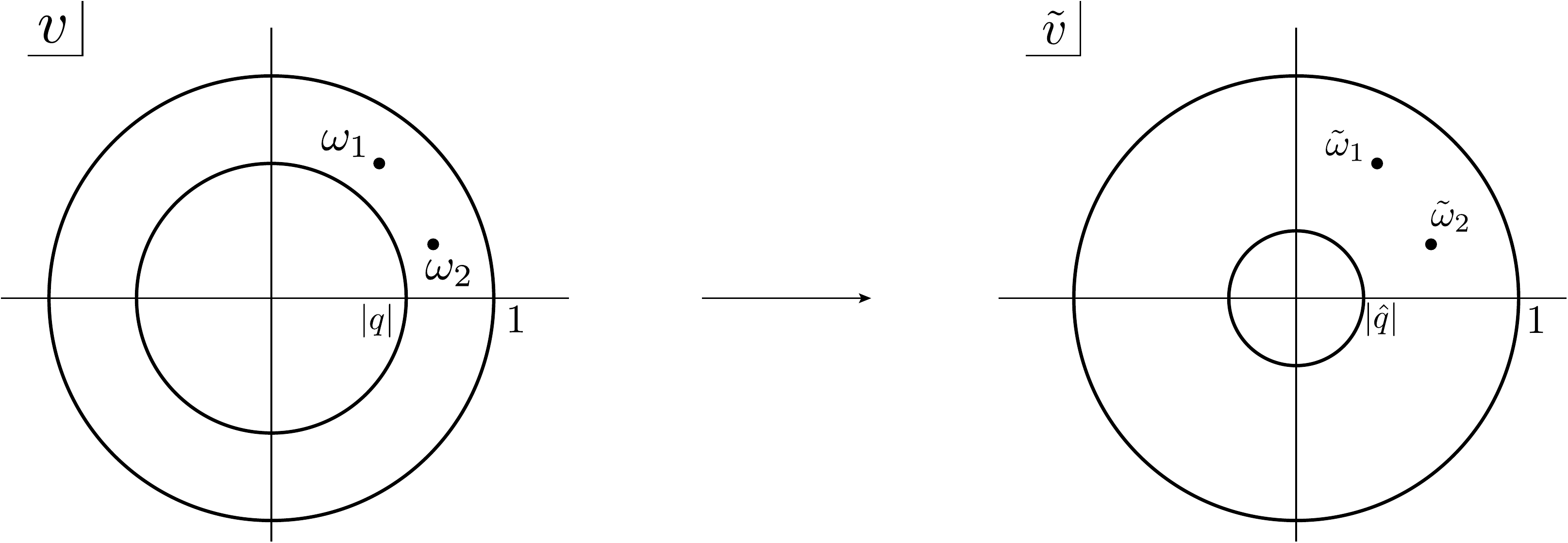}
\caption{Domain of the 2d integral.}
\label{domain}
\end{figure}

The first order correction to the scalar two-point function is given by the integral
\begin{eqnarray}
& & i \frac{\mu}{\pi} \int d^2 v \frac{v^2}{\bar{v}}  
\, w_1^h \, \bar{w}_1^{h} \, w_2^h \, \bar{w}_2^{h} \, 
{\rm Tr}\bigl( W(v) 
\bar{\phi} (w_1,\bar{w}_1) {\phi} (w_2,\bar{w}_2)  \, 
\q^{L_0-\frac{c}{24}} \qb^{\bar{L}_0-\frac{c}{24}} \bigr) \notag \\
& &  \qquad =  i \frac{\mu}{\pi} \int \frac{d^2 v}{v\bar{v}}  \,
F\Bigl( W(v) \bar{\phi}(w_1,\bar{w}_1) \phi(w_2,\bar{w}_2);\tau,\bar{\tau} \Bigr) \ ,
\label{firstorder}
\end{eqnarray} 
where $F$ denotes again the torus amplitude analogous to equation~(\ref{zero}). We are again interested in the high temperature regime, which can be evaluated by doing an $S$-modular transformation, writing the torus amplitude in terms of $\hat\tau=-1/\tau$, and picking out the leading term. Using the familiar modular transformation properties, see e.g.\ \cite{1203.0015}, this leads to 
\begin{eqnarray}
&& i \, \frac{\mu}{\pi} \int \frac{d^2 v}{v \bar{v}} \tauh^{2h+3} \tbh^{2h} 
F \Bigl(  W(v^\tauh) \bar\phi(w_1^\tauh,\bar{w}_1^{\tbh}) {\phi}(w_2^\tauh,\bar{w}_2^{\tbh});\tauh,\tbh \Bigr) \cr
&&\qquad= -i \, \frac{\alpha}{\pi} \, \tauh^{2h+2} \tbh^{2h}  \int \frac{d^2 \vt}{\vt \vbart}
F \Bigl(  W(\vt) \bar\phi(\wt_1,\wbart_1) {\phi}(\wt_2,\wbart_2);\tauh,\tbh \Bigr) \ ,
\label{int}
\end{eqnarray}
where we have renamed the coupling constant $\alpha = -\mu / \tbh$ following the gravity notation, and tilde variables are defined as
\begin{eqnarray}
\tilde{v} \equiv v^\tauh \ , \qquad \tilde{w}_1 \equiv w_1^\tauh \ , \qquad \tilde{w}_2 \equiv w_2^\tauh \ , 
\label{change}
\end{eqnarray}
and similarly for the barred coordinates. The integration domain changes as in Fig. \ref{domain} after the change of variables (\ref{change}).

Let us briefly sketch the issues involved in computing this quantity. We want to extract the leading contribution for $\hat\tau_2\rightarrow \infty$ while maintaining the periodicity of the correlator in the process. Following \cite{Zhu, 1203.0015}, one proceeds to use recursion relations that turn the integrand into a sum of Weierstrass functions multiplying the scalar fields and their derivatives. These capture the interactions between the $W$ current and the scalar fields while respecting the torus periodicity. We then  integrate the resulting function over the full annulus. 

This last step requires a regularization scheme to deal with colliding operators. We will choose to work with a scheme\footnote{This scheme respects the $\vt \mapsto e^{2\pi i} \vt$ period of the annulus. One can also choose a scheme respecting the other periodicity of the annulus $\vt \mapsto \hat{q} \, \vt$ (see appendix \ref{app:other} for details.). We have confirmed that the result is scheme-independent.\label{footnote11}} \cite{Dijkgraaf:1996iy}, where
\begin{equation}\label{scheme}
\frac{1}{\vbart} = \tilde{\bar{{\partial}}} \ln \vbart = \tilde{\bar{{\partial}}} \ln (\vbart \vt) \ .
\end{equation}
Then we can use Stokes' theorem
\begin{equation}\label{scheme1}
\int_M d^2 \vt \, (\tilde\partial A + \tilde{\bar{{\partial}}} \bar{A}) = \frac{i}{2} \oint_{\partial M} (d\vbart A - d\vt \bar{A})
\end{equation}
to rewrite \eqref{int} as a sum of line integrals. In general, these will involve regular parts, coming from the integration over the annulus boundaries, and singular parts, coming from the points $\vt=\wt_i$ where the operators collide. The latter contours are along small `holes' that have been cut out around the scalar insertion points $w_i$ in order to make the integral well-defined; for these contributions, one can ignore the recursion relations and simply use the OPEs in the integrand of \eqref{int}, as we are regulating short-distance singularities that do not see the topology of the torus. 

Using the recursion relations of \cite{Zhu, 1203.0015} which are briefly recapitulated in appendix \ref{app:other}, it is easy to see that the regular parts of \eqref{int} vanish in the scheme (\ref{scheme}). This is due to the periodicity of Weierstrass functions along the angular cycle of the annulus (see, e.g., appendix A of \cite{1203.0015}). What remains is solely the singular part, and so we are entitled to use the OPEs. 

In the high temperature regime, the leading contribution of the trace comes from the vacuum state (after the modular transformation); using Stokes' theorem, the integral \eqref{int} then simplifies to
\begin{equation}\label{3.14}
-\frac{\alpha}{2 \pi} \, \tauh^{2h+2} \tbh^{2h} \, (\qh \qbh)^{-\frac{c}{24}}\,
\wt_1^h \, \wbart_1^{h} \, \wt_2^h \, \wbart_2^{h} 
\oint_{\rm holes} d\vt \, \vt^2 \ln(\vbart \vt) \left\langle W(\vt) \bar\phi(\wt_1,\wbart_1) {\phi}(\wt_2,\wbart_2) \right\rangle_0 \ ,
\end{equation}
where $\langle \cdots \rangle_0$ denotes the correlation function on the sphere. The contour is along the small holes that have been cut out around the insertion points $\wt_1$ and $\wt_2$. Our job is to compute these integrals.

In order to do so, we only need to understand the poles of the integrand. From the OPE of the $W(\vt)$ field with the scalar field $\bar \phi(\wt_1,\wbart_1)$ we get
\begin{equation}
W(\vt) \bar\phi(\wt_1,\wbart_1) = \frac{(W_0 \bar\phi)(\wt_1,\wbart_1)}{(\vt-\wt_1)^3} 
+ \frac{(W_{-1} \bar\phi)(\wt_1,\wbart_1)}{(\vt-\wt_1)^2} + \frac{(W_{-2} \bar\phi)(\wt_1,\wbart_1)}{\vt-\wt_1} + \cdots
\end{equation}
where, in the large $c$ limit, we have 
\begin{eqnarray}
W_0 \, \bar\phi & \equiv & w_{{\rm f}} \, \bar\phi \notag \\[2pt]
W_{-1} \, \bar\phi &=& \frac{3 w_{\rm f}}{2 h_{\rm f}} L_{-1} \bar\phi = \frac{3 w_{{\rm f}}}{2 h_{{\rm f}}} \partial \bar\phi  \label{wrel} \\
W_{-2} \, \bar\phi &=& \frac{3 w_{\rm f}}{h_{\rm f}(2h_{\rm f}+1)} L_{-1}^2 \bar\phi = \frac{3 w_{{\rm f}}}{h_{{\rm f}}(2h_{{\rm f}}+1)} \partial^2 \bar\phi \ . \notag
\end{eqnarray}
Recall that $\bar\phi$ is the representation corresponding to $(\bar{\rm f};0)$, for which the eigenvalues are (in our conventions)
\begin{equation}\label{eigenvals}
\begin{array}{rcl}
h_{{\rm f}}  &=& {\displaystyle \frac{1}{2} (1 + \lambda)} \\[8pt]
w_{{\rm f}}  &=& {\displaystyle \frac{1}{6} (1 + \lambda) (2 + \lambda) =  \frac{1}{3} h_{\rm f} (2 h_{\rm f} +1) \ .}
\end{array}
\end{equation}
Note that similar statements are true for $(0;{\rm f})$ for which the eigenvalues are obtained from the above by replacing $\lambda \mapsto -\lambda$. Furthermore, the conjugate representations have the same $h$ eigenvalues, and opposite $w$-eigenvalues. 

\smallskip

The contribution from the hole at $\wt_1$ thus leads to\footnote{Notice the contour integral is along the clock-wise direction of the small disk cut around the singularity, thus an extra minus sign is added.}
\begin{eqnarray}
&& i \, \alpha \, \tauh^{2h+2} \tbh^{2h} \, (\qh \qbh)^{-\frac{c}{24}}\,
\wt_1^h \, \wbart_1^{h} \, \wt_2^h \, \wbart_2^{h} 
\Bigl( \Bigl[ \frac{3 w_{{\rm f}}}{2} + \frac{3 w_{{\rm f}}}{2 h_{{\rm f}}} \wt_1 \partial_{\wt_1} \Bigr] \notag \\[2pt]
&& \quad + \ln(\wbart_1 \wt_1) 
\Bigl[ w_{{\rm f}}+\frac{3 w_{{\rm f}}}{h_{{\rm f}}} \wt_1 \partial_{\wt_1} 
+ \frac{3 w_{{\rm f}}}{h_{{\rm f}}(2h_{{\rm f}}+1)} \wt_1^2 \partial_{\wt_1}^2 \Bigr] \Bigr) 
\langle \bar\phi(\wt_1,\wbart_1) {\phi} (\wt_2,\wbart_2) \rangle_0 \ .
\label{res}
\end{eqnarray}
Moving the prefactor $\wt_1^h \, \wbart_1^{\bar{h}} \, \wt_2^h \, \wbart_2^{\bar{h}}$ through the derivatives, this becomes
\begin{equation}
i \, \alpha \, \tauh^{2}\, {\cal D}_1 \, 
\langle \bar \phi(w_1,\bar{w}_1) \phi(w_2,\bar{w}_2) \rangle^{(0)} \ ,
\end{equation}
where the differential operator ${\cal D}_1$ is defined by
\begin{equation}
{\cal D}_1 \equiv \frac{3w_{{\rm f}}}{2h_{{\rm f}}} (\wt_1 \partial_{\wt_1})
+ \ln(\wbart_1 \wt_1) \left( \frac{3w_{{\rm f}}}{h_{{\rm f}}(2h_{{\rm f}}+1)} (\wt_1 \partial_{\wt_1})^2 
- w_{{\rm f}} \frac{h_{{\rm f}}-1}{2h_{{\rm f}}+1} \right) \ .
\label{operator1}
\end{equation}
The contribution from the hole $\wt_2$ differs by an overall minus sign, since in replacing $\bar\phi$ by $\phi$, the relations (\ref{wrel}) change by a minus sign. Putting both contributions together, the first order correction (proportional to $\alpha$) equals
\beq
\langle \bar\phi(\wt_1,\bar{\wt}_1) \phi(\wt_2,\bar{\wt}_2)  \rangle^{(\alpha)} = 
i \, \alpha \, \tauh^2 ({\cal D}_1 - {\cal D}_2) \langle \bar \phi(w_1,\bar{w}_1) \phi(w_2,\bar{w}_2) \rangle^{(0)} \ ,
\eeq
where ${\cal D}_2$ is obtained from ${\cal D}_1$ by replacing $\wt_1$ with $\wt_2$,
and $ \langle \bar \phi(w_1,\bar{w}_1) \phi(w_2,\bar{w}_2) \rangle^{(0)}$ is the function of $\wt_i$ and $\wbart_j$ as defined in \eqref{zerores}. 

Finally we return to parallelogram coordinates using eq.~(\ref{plcyl}), i.e.\ we define
\begin{equation}\label{cyltrans}
\tilde{w}_1 = e^{i\hat\tau z_1} \ , \quad \wt_1 \partial_{\wt_1} = \frac{1}{i\hat\tau} \partial_{z_1} \ , \qquad \hbox{etc.} 
\end{equation}
Then we can use the explicit formula from (\ref{2pt}) to deduce that
\begin{equation}
\frac{\langle \bar\phi(z,\bar{z}) {\phi}(0,0) \rangle^{(\alpha)}}{\langle \bar\phi(z,\bar{z}) {\phi}(0,0) \rangle^{(0)}} =
\alpha \, w_{{\rm f}} \, \tauh^2 \, \frac{-3 \sin (\tauh z) + (\tauh z - \tbh \bar{z}) (2 + \cos (\tauh z)) }{2 \sin^2 \frac{\tauh z}{2}} \ ,
\label{linearres}
\end{equation}
where we have set $z_1 = z$, $z_2 = 0$, for simplicity. With the explicit form of the eigenvalues as given in (\ref{eigenvals}), this then reproduces precisely (\ref{grav1loop}) upon taking into account \eqref{zdef} and \eqref{stransform}.

\subsubsection{The structure of the first order result}\label{sec:first}

Before we proceed we should mention that this first order calculation could have also been done more simply, using the fact that the form of the $3$-point function that appears in (\ref{3.14}) is completely fixed by conformal symmetry. (This is a special feature of the $3$-point function, and hence the following argument only works at first order.) Indeed, since $W$, $\bar\phi$ and $\phi$ are (quasi-)primary fields, we know {\it a priori} that 
\begin{equation}\label{threept}
\left\langle W(\vt) \bar\phi(\wt_1,\wbart_1) {\phi}(\wt_2,\wbart_2) \right\rangle_0
= \frac{{\rm const}}{(\wt_1 - \wt_2)^{2h} (\wbart_1 - \wbart_2)^{2h}}
\Bigl[ \frac{(\wt_1 - \wt_2)}{(\vt-\wt_1) (\vt - \wt_2 ) } \Bigr]^3 \ .
\end{equation}
Furthermore, the constant can be determined by considering a contour integral of $\vt$ around $\wt_1$, say,
\begin{eqnarray}
\left\langle (W_0 \bar \phi)(\wt_1,\wbart_1) {\phi}(\wt_2,\wbart_2) \right\rangle_0  
 & = & \frac{1}{2\pi i}\, \oint_{\wt_1} d\vt \, (\vt-\wt_1)^2 \left\langle W(\vt) \bar\phi(\wt_1,\wbart_1) {\phi}(\wt_2,\wbart_2) \right\rangle_0   \notag \\
& = &  \frac{{\rm const}}{(\wt_1 - \wt_2)^{2h} (\wbart_1 - \wbart_2)^{2h}} \ ,
\end{eqnarray}
i.e. the constant equals $w_{\rm f}$, the $W_0$ eigenvalue of $\bar\phi$ (up to the normalisation factor of the $2$-point function). Inserting \eqref{threept} into \eqref{3.14} then gives the final answer \eqref{linearres} as a simple contour integral, including the correct overall coefficient. 

In fact, this result is true for a scalar field $\phi$ in any representation; in our other derivation we used the special null vector structure of $\phi$, see (\ref{wrel}) and (\ref{eigenvals}), but the result does not actually rely on this. It also generalizes to other cases of nonzero higher spin chemical potentials. These statements have also been checked from the bulk and will be useful in the following.

\subsection{Second order correction}

Next we consider the second order correction for which we need to evaluate the integral
\begin{equation}
-\frac{\mu^2 }{2 \pi^2} \int \frac{d^2 v_1}{v_1 \bar{v}_1} \int \frac{d^2 v_2}{v_2 \bar{v}_2} \, 
F\Bigl( W(v_1) \, W(v_2) \, \bar\phi(w_1,\bar{w}_1) \, {\phi}(w_2,\bar{w}_2) ; \tau,\bar{\tau} \Bigr) \ ,
\label{secondorder}
\end{equation}
where $F$ again denotes the torus amplitude.
After the modular transformation and the redefinition \eqref{change}, the integral becomes
\begin{eqnarray}
& & -\frac{\alpha^2}{2 \pi^2} \tauh^{2h+4} \tbh^{2h} (\qh \qbh)^{-\frac{c}{24}}\,
\wt_1^h \, \wbart_1^{h} \, \wt_2^h \, \wbart_2^{h} \notag \\
& & \qquad \qquad
\int \frac{d^2 \vt_1}{\vbart_1} \int \frac{d^2 \vt_2}{\vbart_2} \, \vt_1^2 \vt_2^2 \
\langle  W(\vt_1)  \, W(\vt_2) \,\bar \phi(\wt_1,\wbart_1) \, {\phi}(\wt_2,\wbart_2)  \rangle_0 \ ,
\end{eqnarray}
plus terms subleading in the high temperature limit. 

As at first order, the regular terms will vanish upon integration by parts and use of the recursion relations, so we can jump right to the use of the OPEs. The $4$-point function in the integrand can be evaluated using the OPEs of $W$ with the various fields,
\begin{equation}
\bcontraction{}{W}{}{W}
WW\bar{\phi}\phi+
\bcontraction{}{W}{W}{\bar\phi}
WW\bar{\phi}\phi+
\bcontraction{}{W}{W\bar\phi}{\phi}
WW\bar{\phi}\phi \ .
\label{three}
\end{equation}
The OPEs for the last two terms were already given before, and we denote the contribution of these pieces to the total result as $\langle \bar\phi(z,\bar{z}) {\phi}(0,0) \rangle_W^{(\alpha^2)}$. The OPE of two spin-3 fields takes the form
\begin{align}\label{WWOPE}
W(\vt_1) W(\vt_2) \sim &  \frac{5 c N_3}{6 \, \vt_{12}^6} + 5 N_3 \left[ \frac{L}{\vt_{12}^4}  
+ \frac{1}{2} \frac{L'}{\vt_{12}^3} + \frac{3}{20} \frac{L''}{\vt_{12}^2} + \frac{1}{30} \frac{L'''}{\vt_{12}} \right] \cr
& +\frac{4}{\vt_{12}^2} \left[ U+ \frac{20 N_3}{5c +22} \Lambda^{(4)} \right]
+\frac{2}{\vt_{12}} \left[ U'+ \frac{20 N_3}{5c +22} \partial \Lambda^{(4)} \right] \ ,
\end{align}
where $N_3$ fixes the normalization of the spin-3 current, and our conventions are
\begin{equation}
N_3 = -\frac{1}{5}(\lambda^2-4) \ .
\end{equation}
The composite field $\Lambda^{(4)}$ in (\ref{WWOPE}) will not contribute in the large $c$ limit at O($\alpha^2)$. The first term in (\ref{WWOPE}) is a disconnected diagram which does not contribute in our 
regularization scheme. Therefore, from the contraction of the two $W$ currents we only need to consider the terms involving $U$ and $L$ (and their derivatives). In total, the $O(\a^2)$ correction is thus the sum of three types of terms, which we denote
\beq
\langle \bar\phi(z,\bar{z}) {\phi}(0,0) \rangle^{(\alpha^2)} = \langle \bar\phi(z,\bar{z}) {\phi}(0,0) \rangle_L^{(\alpha^2)}+\langle \bar\phi(z,\bar{z}) {\phi}(0,0) \rangle_U^{(\alpha^2)}+\langle \bar\phi(z,\bar{z}) {\phi}(0,0) \rangle_W^{(\alpha^2)} \ .
\eeq
The $L$ terms (including its derivatives) combine, using integration by parts, to
\begin{eqnarray}
\alpha^2 \tauh^{4}  \frac{N_3}{3}
\Bigl( 2 h_{{\rm f}} + \ln (\wbart_1 \wt_1) (\wt_1 \partial_{\wt_1}) + \ln (\wbart_2 \wt_2) (\wt_2 \partial_{\wt_2}) \Bigr)
\left\langle \bar{\phi}(w_1,\bar{w}_1), {\phi}(w_2,\bar{w}_2)\right\rangle^{(0)} \notag
\end{eqnarray}
thus leading to (after passage to parallelogram coordinates, see eq.~(\ref{cyltrans})) 
\begin{equation}
\frac{\langle \bar\phi(z,\bar{z}) {\phi}(0,0) \rangle_L^{(\alpha^2)}}{\langle \bar\phi(z,\bar{z}) {\phi}(0,0) \rangle^{(0)}} =
\alpha^2 \tauh^4 \, \frac{N_3 h_{{\rm f}} }{3} \Bigl( 2-(\hat\tau z-\hat{\bar{\tau}} \bar{z}) \cot \frac{\hat\tau z}{2} \Bigr) \ .
\label{piece1}
\end{equation}
The $U$ term and its derivative combine to 
\begin{equation}\label{334}
- \frac{\alpha^2}{2 \pi i} \, \tauh^{2h+4} \tbh^{2h} (\hat{q} \hat{\bar{q}})^{-\frac{c}{24}} \wt_1^h \, \wbart_1^{h} \, \wt_2^h \, \wbart_2^{h} \, 
\oint d\vt \, \vt^3 \ln(\vbart \vt) \left\langle U(\vt) \bar\phi(\wt_1,\wbart_1) {\phi}(\wt_2,\wbart_2) \right\rangle_0 \ .
\end{equation}
We can evaluate this using the same argument as in Section~\ref{sec:first}:\footnote{We have also checked that this result is correctly reproduced by the OPEs of $U$ with $\bar\phi$ and $\phi$, i.e.\ doing the analogue of the calculation of Section~\ref{sec:forder}.} the 3-point function is of the form 
\beq
\left\langle U(\vt) \bar\phi(\wt_1,\wbart_1) {\phi}(\wt_2,\wbart_2) \right\rangle_0 = u_{\rm f}\,
\Bigl[ \frac{(\wt_1 - \wt_2)}{(\vt-\wt_1) (\vt - \wt_2 ) } \Bigr]^4\left\langle \bar\phi(\wt_1,\wbart_1) {\phi}(\wt_2,\wbart_2) 
\right\rangle_0  \ ,
\eeq
where $u_{\rm f}$ is the spin-4 zero mode eigenvalue of $\bar\phi$,
\begin{equation}
u_{{\rm f}}  = \frac{1}{20} (1+\lambda)(2+\lambda)(3+\lambda) = \frac{1}{5} h_{\rm f} (2 h_{\rm f} +1)(h_{\rm f}+1) \ .
\label{ueigen} 
\end{equation}
Inserting this into \eqref{334} the subsequent contour integration yields (after passage to parallelogram coordinates)
\begin{equation}
\frac{\langle \bar\phi(z,\bar{z}) {\phi} (0,0) \rangle_U^{(\alpha^2)}}{\langle\bar \phi (z,\bar{z}) {\phi} (0,0) \rangle^{(0)}} =
u_{{\rm f}} \, \alpha^2 \, \tauh^4 \left\{ \frac{19 + 11 \cos (\tauh z)}{6 \sin^2 \frac{\tauh z}{2}} 
- (\tauh z - \tbh \bar{z})\,  \frac{(4 + \cos (\tauh z)) \sin (\tauh z)}{4 \sin^4 \frac{\tauh z}{2}} \right\} \ .
\label{piece2}
\end{equation}

\smallskip

\noindent The second and third terms of \eqref{three}, on the other hand, lead to 
\begin{eqnarray}
-\frac{\alpha^2}{2} \tauh^{4} \Bigl\{
{\cal D}_1 {\cal D}_1 - 2 {\cal D}_1 {\cal D}_2 + {\cal D}_2 {\cal D}_2 \Bigr\}
\left\langle \bar\phi(w_1,\bar{w}_1), {\phi}(w_2,\bar{w}_2)\right\rangle^{(0)} \ ,
\end{eqnarray}
where ${\cal D}_1$ is defined in (\ref{operator1}).
This leads to the contribution
\begin{equation}
\frac{\langle \bar\phi(z,\bar{z}) {\phi}(0,0) \rangle_{W}^{(\alpha^2)}}{\langle \bar\phi (z,\bar{z}) {\phi} (0,0) \rangle^{(0)}} =
\frac{w_{{\rm f}}^2}{16 h_{{\rm f}} (2h_{{\rm f}}+1)} \alpha^2 \tauh^4 
\Bigl[ {\cal N}_0 - (\tauh z-\tbh \bar{z})\, {\cal N}_1 + (\tauh z-\tbh \bar{z})^2 {\cal N}_2 \Bigr] \ , 
\label{piece3}
\end{equation}
with the definitions 
\begin{eqnarray}
{\cal N}_0 & \equiv & \frac{12(2 h_{{\rm f}}+1) ((3 h_{{\rm f}}+1) \cos (\tauh z)+3 h_{{\rm f}}+5)}{\sin^2 \frac{\tauh z}{2}} \notag \\[2pt]
{\cal N}_1 & \equiv & \frac{24(h_{{\rm f}}+1) (h_{{\rm f}} \cos (\tauh z)+2 h_{{\rm f}}+3) 
\sin (\tauh z)}{\sin^4 \frac{\tauh z}{2}} \notag \\[2pt]
{\cal N}_2 & \equiv & \frac{(2 h_{{\rm f}}+1) (2 (4 h_{{\rm f}}+9) \cos (\tauh z)
+h_{{\rm f}} \cos (2 \tauh z))+9 (2 h_{{\rm f}}^2+5 h_{{\rm f}}+4)}{\sin^4 \frac{\tauh z}{2}} \ .
\end{eqnarray}
Collecting all three contributions (\ref{piece1}), (\ref{piece2}), and (\ref{piece3}) together, the final result for the second order correction (i.e.\ the term proportional to $\alpha^2$) is then
\begin{eqnarray}\label{secfin}
&& \frac{\langle \bar\phi(z,\bar{z}) {\phi} (0,0) \rangle^{(\alpha^2)}}{\langle \bar\phi (z,\bar{z}) {\phi} (0,0) \rangle^{(0)}} =
\frac{\alpha^2 \tauh^4 (1+\lambda)(2+\lambda)}{48} 
\Biggl\{ \frac{(\lambda+4)[(3\lambda+13) + (3\lambda+5)\cos(\tauh z)]}{\sin^2 \frac{\tauh z}{2}} \notag \\[2pt]
&& \ \quad- (\tauh z - \tbh \bar{z}) \frac{(\lambda+4)[2(\lambda+4) 
+ (\lambda+1)\cos(\tauh z)] \sin (\tauh z) }{\sin^4 \frac{\tauh z}{2}} \\
&& \ \quad + (\tauh z - \tbh \bar{z})^2 \frac{9(\lambda^2+7\lambda+14)+4(2\lambda^2+15\lambda+22)\cos (\tauh z) 
+(\lambda+1)(\lambda+2) \cos (2 \tauh z)}{12 \sin^4 \frac{\tauh z}{2}} \Biggr\}  \ , \nonumber
\end{eqnarray}
where we have used the eigenvalues (\ref{eigenvals}) and (\ref{ueigen}). We have also confirmed that the same result is obtained  using the other regularization scheme explained in appendix \ref{app:other}.

This result holds for arbitrary $\lambda$; for  $\lambda=-3,-4,-5,-6$, where we can do the gravity calculation at arbitrary temperature (see eqs.~(\ref{n3}) -- (\ref{n6})), it matches what we found there after taking into account the S-modular transformation \eqref{stransform}. 

We can also take the zero temperature limit, discussed in section~\ref{sec:chiral}, of the above result and match to the bulk result \eqref{chl}. This corresponds here to taking $\hat{\tau}_2 \rar 0$ for fixed $\mu=-\a \, \hat\taub$. One easily confirms a match through $O(\mu^2)$ for arbitrary $\l$.

\section{A further test: higher scalar representations}\label{sec:higher}

There are various directions in which one could extend these computations. One is to take the dual CFT primary to live in a higher, i.e.\ non-defining, representation. The bulk calculation then uses a master field $C$ living in a representation $\L^+$ of \hsl, whose lowest component is a scalar field with a different mass \cite{1302.6113}; the CFT representation has a different spectrum, and consequently the relevant OPEs between $\phi$ and the higher spin fields are different. In the context of the corresponding ${\cal W}_N$ minimal models, the relevant CFT primary is then described by $(\L^+ ;0)$.

We will focus our attention on the case of the antisymmetric two-box representation, $\L^+={\sm\yng(1,1)}$. For simplicity we also focus on the `semiclassical limit' of the ${\cal W}_N$ minimal models \cite{1205.2472, 1210.8452}, for which $\l=-N$; in this limit the primary state of the $({\sm\yng(1,1)}\ ;0)$ representation, which we denote $\phi_a$, has conformal dimension 
\beq
h({\sm\yng(1,1)}\ ;0) \equiv h_a = 2h_{\rm f} +1 = 2 + \lambda \ . 
\label{eigenh2}
\eeq
Denoting the CFT primary by $\phi_{a}\equiv ({\sm\yng(1,1)}\ ;0)$, our goal is to compute perturbatively
its two-point function in the bulk and on the boundary,
\beq\lab{exp2}
\la \bar\phi_{a}(z,\zb)\phi_{a}(0,0)\ra = \la \bar\phi_{a}(z,\zb)\phi_{a}(0,0)\ra^{(0)}
+ \sum_{n=1}^{\infty} \la \bar\phi_{a}(z,\zb)\phi_{a}(0,0)\ra^{(\a^n)} \ ,
\eeq
where, again,
\beq
\la \bar\phi_{a}(z,\zb)\phi_{a}(0,0)\ra^{(0)} = 
\left(4 \t \taub \sin{z\over 2\tau}\sin{\zb\over 2\taub}\right)^{-2 h_a}  \ .
\eeq
As we shall see these results will also agree, which can be taken as evidence for the duality itself and further confirmation for this multi-trace correlator prescription.

\subsection{The CFT approach}

The CFT calculation is essentially the same, except that the detailed form of the OPEs with the higher spin fields is now different. Recall that the `wedge' character of the $({\rm f};0)$ representation reads
\begin{equation}
\chi_{({\rm f};0)}^{\rm wedge} = q^{h_{\rm f}} \frac{1}{1-q} = q^{h_{\rm f}} (1+q+q^2+q^3+\cdots) \ ,
\label{characterf}
\end{equation}
from which it follows that there is only one independent descendant state (inside the wedge) at each level. On the other hand, for (${\sm\yng(1,1)}\ ;0$) the wedge character equals
\begin{equation}
\chi_{({\sm\yng(1,1)};0)}^{\rm wedge} = q^{h_a} \frac{1}{(1-q)(1-q^2)} 
=  q^{h_a} (1+q+2 q^2+2 q^3 + \cdots ) \ .
\end{equation}
For example, there is only one state at level one, and hence the null-vector at that level must be the same as before, i.e.\
\begin{equation}
W_{-1} \bar\phi_a = \frac{3w_a}{2h_a} L_{-1} \bar\phi_a = \frac{3w_a}{2h_a} \partial \bar\phi_a \ ,
\end{equation}
where $w_a$ and $h_a$ are the $W_0$ and $L_0$-eigenvalues of $\bar\phi_a$, respectively. At level 2 (and 3), however, there is one more independent state inside the wedge. For example, at level $2$, we may take the two independent states to be $W_{-2} \bar\phi_a$ and $L_{-1}^2 \bar\phi_a$. Thus we cannot replace $W_{-2} \bar\phi_a$ in terms of $L_{-1}^2 \bar\phi_a$ using null vectors as in the last equation of (\ref{wrel}), and we need to proceed differently. In fact, what actually matters for the calculation is to determine
\begin{eqnarray}
\langle (W_{-2}\bar \phi_a)(\wt_1,\wbart_1) \, \phi_a(\wt_2,\wbart_2) \rangle_0 & = & 
\left\langle 
\Bigl(\bigl[ W_{-2} - \frac{3w_a}{h_a (2h_a+1)} L_{-1}^2 \bigr] \bar\phi_a\Bigl) (\wt_1,\wbart_1) \,
\phi_a(\wt_2,\wbart_2)  \right\rangle_0 \notag \\
& & \quad + \frac{3w_a}{h_a (2h_a+1)} \partial^2_{\wt_1}
\langle  \bar\phi_a(\wt_1,\wbart_1) \, \phi_a(\wt_2,\wbart_2) \rangle_0 \notag \\[2pt]
& = & \frac{3w_a}{h_a (2h_a+1)} \, \partial^2_{\wt_1}
\langle \bar\phi_a(\wt_1,\wbart_1)\, \phi_a(\wt_2,\wbart_2) \rangle_0 \ , 
\end{eqnarray}
where we have used that $[ W_{-2} - \frac{3w_a}{h_a (2h_a+1)} L_{-1}^2 ] \bar\phi_a$ is quasi-primary (and hence does not contribute in the two-point function with $\phi_a$). At linear order the calculation therefore goes effectively through: the result is \eqref{linearres}, with the replacement of $w_{\rm f}$ by the $W_0$ eigenvalue of $\bar\phi_a$, which is 
\begin{equation}
w(\overline{\sm\yng(1,1)}\ ;0) \equiv w_a =  \frac{1}{3} (2 + \lambda) (4 + \lambda) \ .
\label{eigenw2}
\end{equation}
Again, this result is also immediately implied by the considerations of Section~\ref{sec:first}.

\smallskip

At second order, the calculation of the $L$ and $U$ contributions, i.e.\ the first term in \eqref{three}, is essentially unchanged --- indeed, effectively this is again a 3-point function calculation, and hence the comments of Section~\ref{sec:first} apply --- but the determination of the second and third terms in \eqref{three} require more care. Since there is no null-vector equation for $W_{-2}\bar\phi_a$, we have to use the usual mode bouncing tricks in order to evaluate these correlators. Apart from that, however, the calculation is very similar, and the final result to second order equals
\begin{eqnarray}
&& \hspace*{-0.2cm} 
\frac{\langle \bar\phi_a(z,\bar{z}) {\phi}_a (0,0) \rangle^{(\alpha^2)}}{\langle \bar\phi_a (z,\bar{z}) {\phi}_a (0,0) \rangle^{(0)}} =
\frac{\alpha^2 \tauh^4 (\lambda+2)}{12} \times \notag \\[2pt]
&& \ \Biggl\{ \frac{1}{\sin^2 \frac{\tauh z}{2}} \bigl[ (3\lambda^3+38\lambda^2+174\lambda+259) 
+(\lambda+5) (3\lambda^2+19\lambda+31)\cos(\tauh z) \bigr] \cr
&& \quad -\frac{(\tauh z - \tbh \bar{z})}{\sin^4 \frac{\tauh z}{2}} \bigl[ (2 \lambda^3+25 \lambda^2+112 \lambda+164)
+(\lambda^3+11 \lambda^2+38 \lambda+43)\cos(\tauh z) \bigr] \sin (\tauh z) \cr
&& \quad + \frac{(\tauh z - \tbh \bar{z})^2}{12 \sin^4 \frac{\tauh z}{2}} 
\bigl[ 9 (\lambda^3+12 \lambda^2+52 \lambda+74)
+ 2(\lambda+4)^2 (4\lambda+17) \cos (\tauh z) \notag \\
& & \hspace{3in} + (\lambda+4)^2 (\lambda+2) \cos (2\tauh z) \bigr] \Biggr\}  \ , \label{2box2}
\end{eqnarray}
where we have used the eigenvalues (\ref{eigenh2}) and (\ref{eigenw2}) and
\begin{equation}
u(\overline{\sm\yng(1,1)}\ ;0) \equiv u_a =  \frac{1}{10} (2+\lambda)(3+\lambda)(7+\lambda)  \ . 
\end{equation}
As before, this result is valid for arbitrary $\lambda$. We should mention that this calculation can be done equally easily in the 't Hooft limit, where the relevant representation is $(\overline{\sm \yng(2)}\ ;0)$, whose $h_a$, $w_a$, $u_a$ eigenvalues are the same as above \cite{Gaberdiel:2011zw}.

\subsection{The gravity approach}

In the gravity theory we can carry out the computations at integer $\l=-N$. The prescription for computing the two-point correlator of the $({\sm\yng(1,1)}\ ;0)$ operator was described in Section~\ref{sec:bulk1}, and again amounts to evaluating \eqref{hw}, where now $|\hw\ra$ and $|-\hw\ra$ are highest and lowest weight states  of the ${\sm\yng(1,1)}$ representation, respectively.

It is easiest to construct this representation via the antisymmetric tensor product of two ${\sm\yng(1)}$ representations. Then the highest and lowest weight states in ${\sm \yng(1,1)}$ are
\beqa
 |\hw\rangle &=& { |1\rangle |2 \rangle -  |2\rangle |1 \rangle \over \sqrt{2}} \\[2pt]
 |-\hw\rangle &=& { |N\rangle |N-1 \rangle -  |N-1\rangle |N \rangle \over \sqrt{2}}\ ,
\eeqa
and so
\beq\langle -\hw | e^{-\Lambda_0} | \hw\rangle = \langle N  |
e^{-\Lambda_0} | 1\rangle \langle N-1  | e^{-\Lambda_0} | 2\rangle -
\langle N  | e^{-\Lambda_0} | 2\rangle \langle N-1  | e^{-\Lambda_0}
| 1\rangle \ ,
\eeq
where the matrix elements on the right-hand side are taken in the defining representation.

\smallskip

At first order, inferring from results at low-lying values of $N$, we find exactly the same universal structure as in the defining representation, only with a different overall coefficient,
\beq
{\langle \bar\phi_{a}(z,\zb)\phi_{a}(0,0)\rangle^{(\a)}\over \langle \bar\phi_{a}(z,\zb)\phi_{a}(0,0)\rangle^{(0)}}  
= {\a \, w_a\over \t^2} { 3\sin\cz+(2+\cos\cz)(\czb-\cz)
\over 2\sin^2{\cz\over 2}} \ ,
\eeq
where
\beq
w_a  = {(2-N)(4-N)\over 3} \ .
\eeq
This matches the CFT result after taking into account the S-modular transformation. 

\smallskip

At second order, we present results for $N=3,4,5,6$, for which we obtain
\vskip .1 in

\bul $N=3:$
%
\begin{flalign}\label{n32}
\frac{\langle \bar\phi_{a}(z,\zb)\phi_{a}(0,0)\rangle^{(\a^2)}}{\langle \bar\phi_{a}(z,\zb)\phi_{a}(0,0)\rangle^{(0)}} 
= & \frac{\a^2}{36 \t^4 \sin^2 \frac{\cz}{2}} [6+4(\cz-\czb)^2 -(6+(\cz-\czb)^2)\cos\cz \notag \\
&\qquad \qquad \quad -6(\cz-\czb)\sin\cz] \ . &&
\end{flalign}
%
This is the same result as \eqref{n3} because for $\mathfrak{sl}(3)$, ${\sm\yng(1,1)} = \overline{\sm\yng(1)}$. 
\vskip .1 in

\bul $N=4:$
%
\begin{flalign}\label{n42}
\frac{\langle \bar\phi_{a}(z,\zb)\phi_{a}(0,0)\rangle^{(\a^2)}}{\langle \bar\phi_{a}(z,\zb)\phi_{a}(0,0)\rangle^{(0)}} 
&= \frac{\a^2}{8 \t^4 \sin^4 \frac{\cz}{2}} \Big[3(5+2(\cz-\czb)^2)-16\cos\cz+\cos2\cz&\notag\\
&\qquad \qquad \qquad +2(\cz-\czb)(-8\sin\cz+\sin2\cz)\Big]\ .&
\end{flalign}
%
\vskip .1 in

\bul $N=5:$
%
\begin{flalign}\label{n52}
\frac{\langle \bar\phi_{a}(z,\zb)\phi_{a}(0,0)\rangle^{(\a^2)}}{\langle \bar\phi_{a}(z,\zb)\phi_{a}(0,0)\rangle^{(0)}} 
&= \frac{\a^2}{16  \t^4 \sin^4{\cz\over 2}} \Big[72+33(\cz-\czb)^2+2(-36+(\cz-\czb)^2)\cos\cz&\notag\\
&\qquad +(\cz-\czb)^2\cos2\cz-(\cz-\czb)(84\sin\cz-6\sin2\cz)\Big]\ . &
\end{flalign}
%
\vskip .1 in

\bul $N=6:$
%
\begin{flalign}\label{n62}
\frac{\langle \bar\phi_{a}(z,\zb)\phi_{a}(0,0)\rangle^{(\a^2)}}{\langle \bar\phi_{a}(z,\zb)\phi_{a}(0,0)\rangle^{(0)}} 
&= \frac{\a^2}{36  \t^4 \sin^4{\cz\over 2}} \Big[9(35+22(\cz-\czb)^2)+8(-30+7(\cz-\czb)^2)\cos\cz
&\notag\\
& \qquad +(-75+16(\cz-\czb)^2)\cos2\cz-30(\cz-\czb)(16\sin\cz+\sin2\cz)\Big] \ .&
\end{flalign}
%

These match the CFT result (\ref{2box2}) for $\lambda=-3,-4,-5,-6$ after taking into account the S-modular transformation.

\section{Discussion}\label{sec:discussion}

The preceding match between CFT scalar correlators as computed from the bulk and the boundary is another piece of evidence for the proposed higher spin AdS$_3$/CFT$_2$ dualities of \cite{Gaberdiel:2010pz, 1205.2472}. Our calculations can be straightforwardly extended in a number of directions, for instance to the supersymmetric realm \cite{Creutzig:2011fe,Tan:2012xi, Chen:2013oxa, Datta:2013qja}.

One issue that has emerged in the course of this investigation is the question of what the correct prescription for the CFT deformation should be. As we have argued here, the deformation should be given by  \eqref{actionpert}. While this is very natural for various reasons, it is not obviously the same as the prescription that was used for the successful match of the black hole and boundary entropies in  \cite{1108.2567,1203.0015}. (In particular, while here the perturbation is via a 2d integral, the perturbative term in  \cite{1108.2567,1203.0015} was taken to be just the zero mode $W_0$, and at least on the face of it, these two descriptions do not agree. See the discussion in Appendix~\ref{app:def}.) A full understanding of this subtlety may also help to cast some light on the recent interpretational mysteries surrounding the higher spin black hole entropy, see \cite{Perez:2013xi, 1302.0816, 1302.1583, gc}.


To close, let us take this opportunity to highlight a pressing open question for the higher spin duality enterprise at large. As we noted, our calculations are driven by the interplay between higher spin symmetry and charged scalar primary operators, and are independent of the specific CFT in question as long as it has the right spectrum. The large $c$ limits of the ${\cal W}_N$ minimal models have this property, but our results do not probe the interactions among the scalar operators themselves. Indeed, in the context of the duality, all explicit higher spin gravity calculations involving scalar fields have so far been at linearized order around a given higher spin background. While this nevertheless allows one to calculate holographically certain three-point and even four-point functions of the dual CFT \cite{1106.2580, 1111.3926, 1302.6113}, these cannot involve more than two light scalar operators. 

Thus, to come closer to a smoking gun for the proposed dualities, we feel that it is important to understand the dynamics on the gravity side beyond the free field level. With respect to calculating correlation functions, one would like to effectively perform a Witten diagram expansion in the bulk. There are plenty of explicit predictions from CFT --- for instance, from four-point functions of scalar primary operators that remain light in the classical limit --- that the bulk theory must reproduce if the duality is valid. 

Understanding the bulk perturbation theory of the 3D Vasiliev theory beyond linearized order would also permit one to address the issue of back-reaction. Can we form a black hole from a collapsing shell of matter? How does this mesh with input from CFT \cite{Chang:2011vka}? In a sense, it is not yet clear whether the integrability of the ${\cal W}_N$ minimal models, or the fact that they are interacting, will determine whether black hole formation is possible.

\acknowledgments

We wish to thank Geoffrey Comp\`ere, Wei Song, Hai-Siong Tan, and especially Tom Hartman and Per Kraus for very helpful conversations. The work of MRG and KJ is supported in part by the Swiss National Science Foundation. EP wishes to thank the ETH Zurich, the Solvay Workshop on `Higher Spin Gauge Theories' in Brussels, the `Holography and its Applications' workshop at the Mitchell Institute at Texas A\&M University, the Institute for Theoretical Physics, University of Amsterdam and the Centro de Ciencias de Benasque Pedro Pascual for hospitality during the course of this work. He has also received funding from the European Research Council under the European Union's Seventh Framework Programme (FP7/2007-2013), ERC Grant agreement STG 279943, “Strongly Coupled Systems”. We all wish to thank the `Higher Spins, Strings and Duality' workshop at the GGI in Florence for hospitality during the intermediate stages of this work.

\appendix 

\section{Details of bulk calculations}
\label{hsl}

\subsection{\hsl\ conventions}
\label{hsconvections}

We follow the conventions of \cite{1209.4937} which we briefly recapitulate here. In the defining representation, our \hsl\ generators $V^s_m$ --- labeled by a spin index, $s=2,3,\ldots$ and a mode index, $|m|<s,$ with $m\in \mathbb{Z}$ --- are built from the $\mathfrak{sl}(2)$ subalgebra as
\beq
V^s_m = (-1)^{s-1-m}{(s+m-1)!\over(2s-2)!} \,
 \Big[ \underbrace{V^2_{-1}, \dots [V^2_{-1}, [V^2_{-1}}_{s-m-1~\rm{terms}}, (V^2_1)^{s-1}]]\Big] \ .
\eeq
At $\l=\pm N$, this also defines our basis of $\mathfrak{sl}(N)$ matrices in the $N\times N$ representation upon removing all generators with $s>N$.

In the defining representation of \hsl, the $\star$ operation is the lone star product \cite{Pope:1989sr}, an infinite-dimensional generalization of $N\times N$ matrix multiplication: that is, the star product can be decomposed as a linear combination of \hsl\ generators, plus an identity element $V^1_0$. The \hsl\ commutator is the star commutator. Explicitly, the lone star product is 
\beq
V^s_m \star V^t_n \equiv \half \sum_{u=1,2,3,...}^{s+t-|s-t|-1}g^{st}_u(m,n;\l)V^{s+t-u}_{m+n}\ ,
\eeq
with structure constants
\beq
g_u^{st}(m,n;\lambda) = {q^{u-2}\over2(u-1)!}\phi_u^{st}(\lambda)N_u^{st}(m,n) \ ,
\eeq
where
\beqa
N_u^{st}(m,n) &&= \sum_{k=0}^{u-1}(-1)^k
\left(\begin{matrix}
u-1 \cr k
\end{matrix}\right)
[s-1+m]_{u-1-k}[s-1-m]_k[t-1+n]_k[t-1-n]_{u-1-k}\notag\\
\phi_u^{st}(\lambda) &&= \ _4F_3\left[\begin{matrix}\half + \lambda ~,~   \half - \lambda  ~,~ {2-u\over 2} ~ ,~ {1-u\over 2}\cr
{3\over 2}-s ~ , ~~ {3\over 2} -t~ ,~~ \half + s+t-u\end{matrix}\Bigg|1\right] \ .
\eeqa
We make use of the descending Pochhammer symbol, $[a]_n  = a(a-1)...(a-n+1)$. $q$ is a normalization constant which we set to $q=1/4$. These structure constants are polynomials in $\l^2$. 
In defining our trace operation, we append no overall normalization factor, i.e.
\beq
\Tr(X) = X|_{V^1_0} \ .
\eeq
An explicit formula for the bilinear trace $\Tr(V^s_mV^s_{-m})$ can be found in, e.g. \cite{1101.2910}.

\subsection{First order perturbation theory}
\label{bulkpt}

Recall the scalar equation \eqref{sc} expanded to first-order in $\a$, given in \eqref{pt1}
\beq\lab{pt2}
d\ch + A\star \ch - \ch\star \Ab = -\Ah \star C
\eeq
and denote $S\equiv -\Ah \star C$. The ingredients are as follows: 
\vskip .1 in
\bul $(A,\Ab)$ are the pure BTZ connections,
\beqa\lab{btzc}
A &&= \left(e^{\rho}V^2_1 +{e^{-\rho}\over 4\t^2}V^2_{-1}\right)dz + V^2_0 d\rho\notag\\
\Ab &&= \left(e^{\rho}V^2_{-1} +{e^{-\rho}\over 4\taub^2}V^2_{1}\right)d\zb - V^2_0 d\rho\ ;
\eeqa

\bul $C$ is the master field in the pure BTZ background obeying $dC + A\star C -C\star \Ab=0$\ ; 

\bul $\Ah$ is the $O(\a)$ part of the \hsl\ black hole connection \eqref{pgf} with charges \eqref{charges}, gauge-transformed to restore $\rho$-dependence using \eqref{CSgaugefields}; explicitly, 
\beq\lab{ahat}
\Ah = {e^{-2\rho}\over 6\t^5}V^3_{-2} dz -{1\over \taub}\left(e^{2\rho}V^3_2 
+ {1\over 16\t^4}V^3_{-2}+{e^{-2\rho}\over 2\t^2}V^3_0\right)d\zb\ ;
\eeq

\bul $\ch$ is the $O(\a)$ fluctuation whose identity component $\phih = \Tr(\ch)$ gives the scalar field perturbation. 

\vskip .1 in

The equation \eqref{pt2} decomposes along spacetime and internal directions to give an infinite set of component equations. We want to decouple these components to extract the wave equation for $\phih$, which was given in \eqref{wave} in terms of components of $S$. From \eqref{pt2}, we see that this will be a two-derivative equation for $\phih$: the only effect of the higher spin deformation of the connection is to generate a source for a scalar field moving in the BTZ background.

By studying the structure of \eqref{pt2} and \eqref{btzc}, one finds a minimal set of six equations needed to extract the wave equation. Denoting by $V^s_{m,x^{\mu}}$ the component of \eqref{pt2} along the generator $V^s_m$ and the direction $dx^{\mu}$, the set is
\beqa\lab{a9}
V^1_{0,\rho}: \quad &&\p_{\rho}\phih+\ch^2_0\cdot{\l^2-1\over 6}=0\notag\\
V^1_{0,z}: \quad && \p \phih-\ep \ch^2_{-1}\cdot{\l^2-1\over 6}-{\emr\over 4\t^2} \ch^2_1\cdot{\l^2-1\over 6}=S^1_{0,z}\notag\\
V^1_{0,\zb}: \quad &&\pb \phih-\ep \ch^2_1\cdot{\l^2-1\over 6}-{\emr\over 4\taub^2} \ch^2_{-1}\cdot{\l^2-1\over 6}=S^1_{0,\zb}\notag\\
V^2_{1,z}: \quad &&\p \ch^2_1+\ep \phih +{\ep\over 2} \ch^2_0-\ep \ch^3_0\cdot {\l^2-4\over 30}-{\emr\over 4\t^2} \ch^3_2\cdot {\l^2-4\over 5}=S^2_{1,z}\notag\\
 V^2_{1,\zb}: \quad && \pb \ch^2_1 - \ep \ch^3_2\cdot{\l^2-4\over 5}
+{\emr\over 4\taub^2}\left(\phih-\half \ch^2_0-\ch^3_0\cdot {\l^2-4\over 30}\right)=S^2_{1,\zb}\notag\\
V^2_{0,\rho}: \quad &&\p_{\rho}\ch^2_0+2\phih+\ch^3_0\cdot {2(\l^2-4)\over 15}=0 \ .
\eeqa
Thus we have six equations for six components, $\lbrace \phih, \ch^2_0, \ch^2_{\pm 1} \ch^3_{0}, \ch^3_2\rbrace$. For $S=0$, these are the equations in the pure BTZ background for the components of $C$. Decoupling these leads to \eqref{wave}. 

Now we need to compute the source terms. Using \eqref{ahat} and computing $S^s_{m,x^{\mu}}$ along the necessary directions using the lone star product, we find
\beqa\lab{sources}
S^1_{0,z} &&= - {(\l^2-1)(\l^2-4)\over 180\t^5}e^{-2\rho}C^3_2\notag\\
S^1_{0,\zb} &&= -{(\l^2-1)(\l^2-4)\over 1440\taub}\left(48e^{2\rho}C^3_{-2} + {3\over \t^4}e^{-2\rho}C^3_2+{4\over \t^2}C^3_0\right)\notag\\
S^2_{1,z} &&= -{(\l^2-4)(\l^2-9)\over 140\t^5}e^{-2\rho}C^4_3\notag\\
S^2_{1,\zb} &&= -{(\l^2-4)\over \taub}\Bigg[e^{2\rho}\left(-{1\over 5}C^2_{-1}-{1\over 10}C^3_{-1}+{(\l^2-9)\over 350}C^4_{-1}\right)\notag\\&& \quad 
+ {e^{-2\rho}\over \t^4}\left({3(\l^2-9)\over 1120}\right)C^4_3+{1\over \t^2}\left(-{1\over 60}C^2_1-{1\over 40}C^3_1+{(\l^2-9)\over 700}C^4_1\right)\Bigg] \ .
\eeqa
We see that we need to derive the following components of the master field $C$ in the BTZ background: $\lbrace C^3_{\pm2}, C^3_{\pm1}, C^3_{0}, C^4_3, C^4_{\pm1}, C^2_{-1}\rbrace$. In particular we need to solve for these in terms of the scalar field $\Phi^{(0)}\equiv \Tr(C)$, where $\Phi^{(0)}$ is the zeroth order bulk-boundary propagator \eqref{btzprop}. Thus the main task at this stage is to find a closed set of equations for $C$ with which one can solve for the desired components of $C$.

Inspection of the BTZ connection and the nature of the lone star product guides one toward the following (non-unique) algorithm: 

\vskip .1 in

1. Solve, in sequence, $V^1_{0,z}, V^1_{0,\zb}$ for $C^2_{\pm1}$; then $V^2_{0,z}, V^2_{0,\zb}$ for $C^3_{\pm1}$; and finally $V^3_{0,z}, V^3_{0,\zb}$ for $C^4_{\pm1}$.

2. Next, solve $V^1_{0,\rho}, V^2_{0,\rho}$ for $C^2_{0}$ and $C^3_0$.

3. Solve $V^2_{1,z}, V^2_{-1,\zb}$ for $C^3_{\pm 2}$. 

4. Finally, solve $V^3_{2,z}$ for $C^4_3$. 

\vskip .1 in

The logic of some of these steps is manifest in \eqref{a9}, recalling that $C$ obeys those same equations with $S=0$. For more guidance and some mnemonics on how to read off quickly which components of $C$ appear in which equations, we refer the reader to \cite{1111.3926}. 

Continuing, this algorithm yields the sources \eqref{sources}, given in terms of spacetime derivatives of $\Phi^{(0)}$. In particular, the $S^1_0$ terms contain up to two derivatives of $\Phi^{(0)}$, and the $S^2_1$ terms contain up to three derivatives; all are fairly long expressions. Plugging them into \eqref{wave} yields the final wave equation for $\phih$, which is a long result. Nevertheless, it is easy to check that \eqref{prop} is a solution.

\section{Another regularization scheme}\label{app:other}

In this appendix we repeat the first order correction (\ref{firstorder}), using the regularization scheme that preserves the $\vt \mapsto \hat{q} \, \vt$ period of the annulus, as advertised in footnote~\ref{footnote11}. By modular invariance, this is equivalent to using the scheme that preserves the angular periodicity {\it before} the modular transformation. The second order correction (\ref{secondorder}) can be worked out similarly, but is more involved.

In this regularization, before the modular transformation, we write
\begin{equation}
\frac{1}{\bar{v}} = \bar\partial \ln(\bar{v} v) \ .
\end{equation}
Then using Stokes' theorem, (\ref{firstorder}) becomes
\begin{equation}\label{B.2}
\frac{\mu}{2 \pi} \oint \frac{dv}{v} \ln(\bar{v} v)
F\Bigl( W(v) \bar{\phi}(w_1,\bar{w}_1) \phi(w_2,\bar{w}_2);\tau,\bar{\tau} \Bigr) \ ,
\end{equation}
where the contour integral runs along the boundaries (which gives the regular part) and encircles the singularities (which gives the singular part). 

For the regular part, the integration along the outer circle is trivially zero since the radius is $1$ (and hence the logarithm vanishes), while the integration along the inner circle gives
\begin{equation}\label{B.3}
2 \mu \tau_2 \oint_{0} \frac{dv}{v}
F\Bigl( W(v) \bar{\phi}(w_1,\bar{w}_1) \phi(w_2,\bar{w}_2);\tau,\bar{\tau} \Bigr) \ .
\end{equation}
The contour integral isolates the zero mode of $W(v)$, so the calculation is similar to those in \cite{1203.0015}. Following the standard procedure to compute torus amplitudes at high temperature, we now perform a S-modular transformation so that the low-lying states dominate. After a change of variables (\ref{change}), we get
\begin{equation}
-2 \alpha \, \tauh_2 \, \tauh^{2h+1} \tbh^{2h} \int_1^{\qh} \frac{d\vt}{\vt}
F\Bigl( W(\vt) \bar{\phi}(\wt_1,\wbart_1) \phi(\wt_2,\wbart_2);\tauh,\tbh \Bigr) \ .
\end{equation}
This can be calculated using the recursion relations (see (2.21) of \cite{1203.0015}) and we have
\begin{eqnarray}
& {\displaystyle 
-2 \alpha \, \tauh_2 \, \tauh^{2h+1} \tbh^{2h} \int_1^{\qh} \frac{d\vt}{\vt} 
\Bigl\{ \sum_{m\geq 0} {\cal P}_{m+1} \left( \frac{\wt_1}{\vt} \right)
F\Bigl( W[m] \bar{\phi}(\wt_1,\wbart_1) \phi(\wt_2,\wbart_2);\tauh,\tbh \Bigr)} \\
& \qquad \qquad \qquad \qquad \qquad \qquad 
{\displaystyle + \sum_{m\geq 0} {\cal P}_{m+1} \left( \frac{\wt_2}{\vt} \right)
F\Bigl( \bar{\phi}(\wt_1,\wbart_1) W[m] \phi(\wt_2,\wbart_2);\tauh,\tbh \Bigr) \Bigr\}\ , } \nonumber
\label{b5}
\end{eqnarray}
where ${\cal P}_{m+1}$ denote the Weierstrass functions, and the bracketed modes are defined the same way 
as in (2.22) of \cite{1203.0015}. Using the integrals of Weierstrass functions
\begin{eqnarray}
\int_1^{\qh} \frac{d\vt}{\vt} {\cal P}_{1} \left( \frac{\wt}{\vt} \right) &=& (2 \pi i)(i \pi - 2 \pi i \tauh + \ln \wt) \\
\int_1^{\qh} \frac{d\vt}{\vt} {\cal P}_{2} \left( \frac{\wt}{\vt} \right) &=& (2 \pi i)^2 \\
\int_1^{\qh} \frac{d\vt}{\vt} {\cal P}_{m+1} \left( \frac{\wt}{\vt} \right) &=& 0 \ , \qquad (m > 1)
\end{eqnarray}
as well as the definition of the bracketed modes
\begin{eqnarray}
W[0] &=& (2 \pi i)^{-1} \left( W_{-2} + 2 W_{-1} + W_0 \right) \\
W[1] &=& (2 \pi i)^{-2} \left( W_{-1} + \frac{3}{2} W_0 + \frac{1}{3} W_1 + \ldots \right)  \ ,
\end{eqnarray}
we find that the first order correction (\ref{B.3}) equals
\begin{equation}
\langle \bar\phi(\wt_1,\bar{\wt}_1) \phi(\wt_2,\bar{\wt}_2)  \rangle_{\rm reg}^{(\alpha)} =  
\left(1 - \frac{\tbh}{\tauh} \right) i \, \alpha \, \tauh^2 \, ({\cal D}_1^{\rm holo} - {\cal D}_2^{\rm holo}) \,
\langle \bar \phi(w_1,\bar{w}_1) \phi(w_2,\bar{w}_2) \rangle^{(0)} \ ,
\end{equation}
where the differential operator
\begin{equation}
{\cal D}_1^{\rm holo} \equiv 
\frac{3w_{{\rm f}}}{2h_{{\rm f}}} (\wt_1 \partial_{\wt_1})
+ \ln(\wt_1) \left( \frac{3w_{{\rm f}}}{h_{{\rm f}}(2h_{{\rm f}}+1)} (\wt_1 \partial_{\wt_1})^2 
- w_{{\rm f}} \frac{h_{{\rm f}}-1}{2h_{{\rm f}}+1} \right) 
\end{equation}
is the `holomorphic' part of (\ref{operator1}). Using the coordinate transformation \eqref{cyltrans}, this leads to the result
\begin{equation}
\frac{\langle \bar\phi(z,\bar{z}) {\phi}(0,0) \rangle_{\rm reg}^{(\alpha)}}{\langle \bar\phi(z,\bar{z}) {\phi}(0,0) \rangle^{(0)}} =
\left(1 - \frac{\tbh}{\tauh} \right) \alpha  \, w_{{\rm f}} \, \tauh^2 \, \frac{-3 \sin (\tauh z) + (\tauh z) (2 + \cos (\tauh z)) }{2 \sin^2 \frac{\tauh z}{2}} \ ,
\label{linearresholo}
\end{equation}
which is the `holomorphic' part of (\ref{linearres}) with an additional factor $(1 - \tbh / \tauh)$.

It remains to determine the contour integral around the singular part of (\ref{B.2}). After the modular S-transformation, this takes the same form as the calculation in section~\ref{sec:forder}, except that we now write instead of 
(\ref{scheme})
\begin{equation}\label{scheme2}
\frac{1}{\vbart} = \tilde{\bar{\partial}} \left[ \ln \vbart  + \frac{\tbh}{\tauh} \ln \vt \right] \ ,
\end{equation}
which is invariant under $\vt \mapsto \qh \vt$. Relative to \eqref{linearres}, the effect of (\ref{scheme2}) is to produce an overall factor $\tbh / \tauh$ for the `holomorphic' piece (the piece proportional to $\ln \vt$), while keeping the piece proportional to $ \ln \vbart $ unchanged. Thus, the result of the singular part is
\begin{equation}
\frac{\langle \bar\phi(z,\bar{z}) {\phi}(0,0) \rangle_{\rm sing}^{(\alpha)}}{\langle \bar\phi(z,\bar{z}) {\phi}(0,0) \rangle^{(0)}} =
 \alpha  \, w_{{\rm f}} \, \tauh \tbh \, \frac{-3 \sin (\tauh z) + \tauh (z-\bar{z}) (2 + \cos (\tauh z)) }{2 \sin^2 \frac{\tauh z}{2}} \ .
\label{linearresantiholo}
\end{equation}
Adding (\ref{linearresholo}) to (\ref{linearresantiholo}), we then precisely recover (\ref{linearres}).

\section{Black hole partition function from CFT redux}\label{app:def}

As was mentioned at the beginning of Section~\ref{sec:cft}, the perturbation that is used in this paper, eq.~(\ref{actionpert}), differs from the perturbing term in the analysis of the black hole entropy (or partition function) that was employed in \cite{1108.2567, 1203.0015}. In this appendix we want to comment on the relationship between the two approaches.

In \cite{1108.2567, 1203.0015}  the perturbed partition was taken to equal
\beq
Z_{\rm 1d} = \Tr\left(e^{2\pi i \a W_0} \q^{L_0-\frac{c}{24}} \, \qb^{\bar{L}_0-\frac{c}{24}} \right) \ ,
\eeq
where $W_0$ means that the zero mode of $W$ has been inserted into the trace. 
This is a natural definition from the perspective of the partition function as a trace over states in 
a Hilbert space. 
At $O(\a^2)$, for example, one must then calculate
\begin{eqnarray}\label{cal1}
Z_{\rm 1d}^{(2)} &=& \frac{(2\pi i \alpha)^2}{2!} \Tr( W_0 W_0 \, q^{L_0-\frac{c}{24}} \, \qb^{\bar{L}_0-\frac{c}{24}}) \cr
&=& \frac{\alpha^2}{2} \oint_0 dv_1 \oint_0 dv_2 \, v_1^2 v_2^2 \, \Tr( W(v_1) W(v_2) q^{L_0-\frac{c}{24}} \, \qb^{\bar{L}_0-\frac{c}{24}}) \cr
&=& \frac{\alpha^2}{2} \oint_0 \frac{dv_1}{v_1} \oint_0 \frac{dv_2}{v_2} \, F(W(v_1) W(v_2);\tau,\taub) \ ,
\end{eqnarray}
where $F$ is the torus amplitude defined as in (\ref{zero}), and the contour encircles the origin.

This is to be compared with what the perturbation (\ref{actionpert}) would give rise to. Then we would
define the deformed partition function as 
\begin{equation}\label{2d}
Z_{\rm 2d}=
\Tr \Bigl(
e^{i\frac{\mu}{\pi} \int d^2 v   \frac{v^2}{\bar{v}} W(v) } 
\q^{L_0-\frac{c}{24}} \, \qb^{\bar{L}_0-\frac{c}{24}} \Bigr) \ ,
\end{equation}
and the second order term is
\begin{equation}\label{cal2}
Z_{\rm 2d}^{(2)}=-\frac{\mu^2}{2 \pi^2} \int \frac{d^2 v_1}{v_1 \bar{v}_1} \int \frac{d^2 v_2}{v_2 \bar{v}_2} \, F(W(v_1) W(v_2); \tau,\taub)  \ .
\end{equation}
We want to study the question whether the two deformations give the same partition function, i.e. whether 
$Z_{\rm 2d} = Z_{\rm 1d}$. 

Formally we may relate the two calculations by simply performing the integral of the perturbing term in
the exponent. Using the mode expansion  $W(v) = \sum_{n} v^{-n-3} W_n$, we then find
\begin{eqnarray}
&& i \frac{\mu}{\pi} \int d^2 v \frac{v^2}{\bar{v}} W(v) = i \frac{\mu}{\pi} \int \frac{d^2 v}{v \bar{v}} \sum_n v^{-n} \, W_n \cr
&&\qquad = i \frac{\mu}{\pi} \int_{\vert q \vert}^1 \frac{dr}{r} \int_0^{2 \pi} d\theta \sum_n r^{-n} e^{-i n \theta} \, W_n
= 4 \pi i \, \mu \, \tau_2 W_0 \ . \label{C.5}
\end{eqnarray}
This suggests that  the partition function (\ref{cal2}) reduces to (\ref{cal1}) provided we identify the chemical potentials as 
\begin{equation}
2 \pi i \, \alpha_{\rm new} = 4 \pi i \, \mu \tau_2 \qquad \Longrightarrow \qquad \alpha_{\rm new} = 2 \mu \tau_2 \ . 
\end{equation}
Note that this differs from the relation between $\mu$ and $\alpha$ that was used in \cite{1108.2567, 1203.0015} as well as earlier in this paper, see the comment after eq.~(\ref{int}),  
\begin{equation}
\alpha =  \mu \bar{\tau} \ . 
\end{equation}
Note that this discrepancy is exactly of the same type as the relation between the holomorphic and canoncial approaches to the calculation of the black hole entropy, see eq.\ (5.30) of \cite{1302.0816}. Thus we suspect that this is related to the inherent ambiguities in defining thermodynamic variables for these theories.\footnote{We thank Tom Hartman for drawing our attention to this.}

\smallskip

One may be worried about the somewhat formal argument in eq.~(\ref{C.5}) above since it does 
not address possible singularities when the operators coincide. For a more rigorous treatment, we may 
use the result of \cite{Dijkgraaf:1996iy} where it is shown that the two approaches only differ by a contact term. At second order, one finds, see \cite[eq.\ (3.14)]{Dijkgraaf:1996iy}
\begin{equation}\label{dijk1}
\left\langle \int W \int W \right\rangle =(2 \pi i \t_2)^2\left\langle \oint W \oint W \right\rangle + 2 \pi^2 i \t_2 \left\langle \oint (WW)_2 \right\rangle \ .
\end{equation}
Here we have used the shorthand notation
\begin{equation}\label{dijk2}
\int W = \int \frac{d^2 v}{v \bar{v}} W(v) \ , \qquad \oint W = \oint_0 \frac{dv}{v} W(v) \ ,
\end{equation}
(with the understanding that they appear inside a torus amplitude) and $(WW)_2$ is the coefficient of the OPE
\begin{equation}
W(v) \cdot W(w) \sim \sum_{n=-\infty}^\infty (v-w)^{-n} \, (WW)_n (w) \ .
\end{equation}
This was derived by using the same scheme as in appendix \ref{app:other}, recursively eliminating all 2d integrals in favor of contour integrals: the first term of the right side of \eqref{dijk1} captures the regular part of the left hand side integrated along the annulus boundary, while the second term comes from the singularities.  Since the ${\cal W}_{\infty}[\mu]$ algebra is `abelian' in the sense of \cite{Dijkgraaf:1996iy} --- this is basically a direct consequence of the fact that the algebra respects a $\mathbb{Z}_2$-grading, where fields of even (odd) conformal dimension are even (odd) --- the two-product $(AB)_2$ does not contain any central terms.\footnote{Note that the ${\cal W}_{\infty}[\mu]$ algebra only contains fields of conformal dimension $h\geq 2$.} Thus, the contact term is subleading in the large central charge limit, and hence does not contribute to the calculation, and we conclude that (\ref{dijk1}) establishes the desired relation.


We should stress though that these manipulations are very subtle. For example, if one uses \eqref{dijk1} after the modular transformation instead --- that is, at low temperature --- the result naively appears to vanish. Similarly, one can use a regularization scheme in which the thermal periodicity $v\mapsto qv$ is preserved instead, in which case its application at {\it high} temperature appears to give zero. (This `flip' is natural, as the modular transformation exchanges cycles.) This seems related to the fact that the respective cycles are becoming degenerate  in the corresponding limits. The scalar calculations suffer no such apparent scheme-dependence, because one picks up singularities from the scalar operator insertions on the torus upon taking the limit. 

In addition, it seems nontrivial to firmly establish that one never picks up $O(c)$ contact terms at higher orders, in the high temperature regime. For instance, one finds terms proportional to the vacuum expectation value of a pair of two-products $(WW)_2$, which will include stress tensor two-point functions. One can provide arguments that such terms vanish at high temperature even if they don't vanish identically, but we leave a full investigation for the future.

\end{document}